\documentclass[aps,pre,epsf,superscriptaddress,amsmath,amssymb,amsfonts,twocolumn,showpacs]{revtex4-1}
\usepackage{amsfonts}
\usepackage{amsmath}
\DeclareMathOperator{\csch}{csch}

\usepackage{amssymb}
\usepackage{graphicx}
\usepackage{dcolumn}
\usepackage{times}
\usepackage{color}
\usepackage{appendix}
\begin{document}
  
\title{Dark-bright soliton interactions beyond the integrable limit}

\author{G. C. Katsimiga}
\email{lkatsimi@physnet.uni-hamburg.de}
\affiliation{Zentrum f\"ur Optische Quantentechnologien,
Universit\"at Hamburg, Luruper Chaussee 149, 22761 Hamburg, Germany}

\author{J. Stockhofe}
\affiliation{Zentrum f\"ur Optische Quantentechnologien,
Universit\"at Hamburg, Luruper Chaussee 149, 22761 Hamburg, Germany}

\author{P. G. Kevrekidis}
\affiliation{Department of Mathematics and Statistics, University of Massachusetts,
Amherst, MA 01003-4515, USA}

\author{P. Schmelcher}
\affiliation{Zentrum f\"ur Optische Quantentechnologien,
Universit\"at Hamburg, Luruper Chaussee 149, 22761 Hamburg, Germany}
 \affiliation{The Hamburg Centre for Ultrafast Imaging,
Luruper Chaussee 149, 22761 Hamburg, Germany}

\begin{abstract}
In this work we present a systematic theoretical analysis regarding dark-bright solitons and their interactions, 
motivated by recent advances in atomic two-component repulsively interacting Bose-Einstein condensates.
In particular, we study analytically via a two-soliton ansatz adopted
within a variational formulation
the interaction between two dark-bright solitons in a homogeneous
environment beyond the integrable regime, by considering general inter/intra-atomic interaction coefficients.
We retrieve the possibility of a fixed point in the case where the bright solitons are out of phase. 
As the inter-component interaction is increased,
we also identify an exponential instability of the two-soliton state, associated
with a subcritical pitchfork bifurcation. The latter gives rise
to an asymmetric partition of the bright soliton mass and dynamically
leads to spontaneous splitting of the bound pair. In the case of
the in-phase bright solitons, we explain via parsing the
analytical approximations and monitoring the direct dynamics
why no such pair is identified,
despite its prediction by the variational analysis.
\end{abstract}

\maketitle

\section{Introduction}

Over the past decade, multi-component Bose-Einstein condensates (BECs)
have offered a fertile ground for the examination of nonlinear
wave phenomena~\cite{revip}. Such systems and their solitary
waves were explored earlier in settings of nonlinear optics~\cite{yuri1}.
There, structures such as the dark-bright (DB) solitons
have been  ubiquitously identified in two-component systems
featuring self- and cross-repulsion (or self- and cross-phase
modulation). As a result, these solitary wave states were extensively
studied~\cite{christo,vdbysk1,vddyuri,ralak,dbysk2,shepkiv,parkshin},
and pioneering experiments featuring individual DBs,
as well as molecules thereof were performed~\cite{seg1,seg2}.
In the far more recent atomic realm, the possibility of
exploring in ultracold gases different hyperfine states of,
e.g., $^{87}$Rb and $^{23}$Na, has created a new and extremely
controllable venue for identifying and revealing the dynamics
of such DB states. Following the theoretical proposal
of~\cite{BA}, recent experiments have, thus, examined
systematically the interaction of DB solitons with
each other, as well as the interplay with external traps~\cite{hamburg,pe1,pe2,pe3,azu}.
Additionally, variants thereof involving SO$(2)$ rotation in the
form of dark-dark solitons have also been monitored~\cite{pe4,pe5}.

In this context, the examination of DB soliton interactions is
an especially intriguing topic. For dark one-component solitons,
the interaction effect for (local) cubic nonlinearities is
one of repulsion~\cite{krolkiv,markus1}. On the other hand,
for bright solitons, the effect is crucially dependent on their relative
phase, as has been recently experimentally also illustrated in
the atomic realm in~\cite{borisrh} for attractive condensates.
DB solitons bear both of these features and additionally
exhibit interactions of the dark solitons of one component with bright
ones of the other, an interaction, arguably, less explored.
The theoretical and experimental work of~\cite{seg2} already
formulated one of the most important pieces of the relevant
intuition: namely, while dark solitons repel, bright ones in
self-defocusing media will attract if they are out-of-phase
and repel if they are in-phase, oppositely to what is the case
for self-attractive/focusing media. Hence, the combination
of repulsion, mediated by the dark solitons
at short distances, and attraction, mediated by the
bright solitons at longer
distances (see below regarding the different range of the
interactions), should lead to an effective two-soliton equilibrium.
Thus, this mechanism may create
a genuine and potentially robust bound molecule
consisting of two DB solitons. This possibility was further
explored in the context of atomic BECs in~\cite{pe3} also including
the effect of a parabolic trap.

However, numerous questions still remain. A principal one
explored here is that of the persistence of such a state
under parametric variations. In particular,
motivated by the tunability of interatomic interactions,
by modifying the $s$-wave 
scattering lengths via the well established technique of
Feshbach resonances~\cite{fr1,fr2,chin},
we revisit DB solitons and 
their interactions through
a direct analytical and numerical investigation of their
static and dynamical properties, when the integrability is 
broken due to unequal inter and intra-species interaction coefficients.
As a case example, we fix the intra-species coefficients $g_{11}$ and $g_{22}$
to ratios of interest for  $^{87}$Rb experiments
(i.e., $1$ and $0.95$ respectively~\cite{halll}; see also~\cite{opanchuk})
and vary 
the experimentally accessible inter-component interaction
coefficient $g_{12}$.
This leads us to identify symmetry-breaking
bifurcations of the state with the anti-symmetric bright
component and associated instabilities leading to travelling
and redistribution of the corresponding mass. Another question
concerns the fate of the case where the bright solitons
are in-phase.  For the latter, our numerical computations
(both fixed point and dynamical ones)
strongly suggest that no equilibrium configuration exists.
For both cases, we employ an analytical calculation based on
a variational two-soliton ansatz. Upon exploring the adequacy of
the ansatz, we use it in both the in- and out-of-phase cases.
For the latter, we predict the location of the equilibrium
in good agreement with the full numerical results. The
former also appears to identify an equilibrium, although
the full numerical computation does not support it. We use
this as a cautionary tale about the validity of the conclusions
of the variational approximation. Additional spurious features
resulting from the ansatz are also discussed. Finally, we
complement the analytical investigations with direct numerical
simulations corroborating our existence and stability conclusions.

The paper is organized as follows: in Section II we
present our theoretical considerations. We start by 
introducing the physical model -- a system of coupled Gross-Pitaevskii equations 
(GPEs) -- and describe previous analytical results existing in the literature   
regarding single DB solitons for general inter/intra-atomic species interactions.
Subsequently, we  use the single DB soliton solution as a building block 
in order to consider two DB solitons, and study their static and dynamical properties. 
From the analytical side, an extension of the effective particle picture
put forth in~\cite{pe3} that captures the aforementioned 
properties for multiple DB solitons and arbitrary interatomic interaction
settings is developed. The resulting effective energy landscape is
presented and discussed.
We show results for both in-phase (IP) and out-of-phase (OP) bright soliton components,
and in both the miscible and immiscible regimes. Recall that
the latter refer to the absence or presence of phase separation
depending on whether or not (respectively) the condition
$g_{11} g_{22} > g_{12}^2$ is met~\cite{aochui}. 
In Section III, we test the theoretical predictions by means
of full numerical computations. Stationary states are attempted
to be identified by fixed point iterations. When they are, their
stability is explored by means of Bogolyubov-de Gennes (BdG)
linearization analysis~\cite{revip} to identify the fate
of small perturbations. Then, the fully nonlinear dynamics
is employed in order to reveal the fate chiefly of the unstable
solutions. Finally, in Section IV we summarize our findings and discuss future 
challenges. 

\section{Theoretical Analysis}

\subsection{Basic properties of single dark-bright solitons}

Our system of interest consists of
a two-component elongated (along the $x$-direction) repulsive 
BEC, composed of two different
hyperfine states of the same alkali isotope, such as $^{87}$Rb.
Assuming that the trap is 
highly anisotropic, with the longitudinal and transverse trapping 
frequencies being such that $\omega_x \ll \omega_{\perp}$, we may 
describe this system by the following two coupled GPEs 
\cite{stringari,siambook}:
\begin{eqnarray}
i\hbar \partial_t \psi_j =
\left( -\frac{\hbar^2}{2m} \partial_{x}^2 +V(x) -\mu_j + \sum_{k=1}^2 
g_{jk} |\psi_k|^2\right)\!\psi_j.\nonumber\\
\label{model}
\end{eqnarray}
Here, $\psi_j(x,t)$ ($j=1,2$) denote the mean-field wave functions of the 
two components normalized to the
numbers of atoms $N_j = \int_{-\infty}^{+\infty} |\psi_j|^2 dx$, $m$, and $\mu_j$ 
are the atomic mass and chemical potentials, respectively. 
The effective 1D coupling constants are given by $g_{jk}=2\hbar\omega_{\perp} a_{jk}$, 
where $a_{jk}$ denote the three $s$-wave scattering lengths (note that 
$a_{12}=a_{21}$) that account for collisions between atoms belonging to 
the same ($a_{jj}$) or different ($a_{jk}, j \ne k$) species, and 
$V(x)$ represents the external trapping
potential.

Measuring densities {$|\psi_j|^2$}, 
length, time and energy in
units of $2a_{11}$, $a_{\perp} = \sqrt{\hbar/\left(m \omega_{\perp}\right)}$, 
$\omega_{\perp}^{-1}$ and $\hbar\omega_{\perp}$,
respectively, we may cast Eqs.~(\ref{model}) into the following dimensionless form:
\begin{eqnarray}
i \partial_t u_d  =& -&\frac{1}{2} \partial_{x}^2u_d  + V(x)u_d
+(|u_d|^2 + g_{12}|u_b|^2 -\mu_d) u_d,\nonumber \\
\label{deq1}
\\
i \partial_t u_b  =& -&\frac{1}{2} \partial_{x}^2u_b +V(x)u_b 
+ (g_{12}|u_d|^2 + g_{22}|u_b|^2- \mu_b) u_b.\nonumber \\
\label{deq2}
\end{eqnarray}
In the above equations, 
we have used the notation $\psi_1 \equiv  u_d$ and $\psi_2 \equiv  u_b$, 
indicating that the component $1$ ($2$) will be 
supporting a dark (bright) soliton. Furthermore,  $\mu_j$, $\left(j=d,b\right)$ are the chemical 
potentials that characterize each component, while the interaction coefficients are normalized to the scattering 
length $a_{11}$, that is $g_{12}\equiv g_{12}/g_{11}$, 
and $g_{22}\equiv g_{22}/g_{11}$. 
Upon considering a standard harmonic potential confining the atoms its form 
in dimensionless units is given by:  
$V(x)=(1/2)\Omega^2 x^2$, where the normalized trap strength is
$\Omega = \omega_x/\omega_\perp$,  thus also representing a natural small 
parameter of the system. 
In what follows, to avoid the additional complications
of the trap, we will consider the simplest possible
case of the homogeneous system, setting $\Omega \rightarrow 0$.

Then, rescaling space-time coordinates as $t \rightarrow 
\mu_d t$, $x \rightarrow {\sqrt{\mu_d}}x$,
and the densities $|u_{d,b}|^2 \rightarrow \mu_{d}^{-1} |u_{d,b}|^2$,
the above system of coupled 
GPEs takes the form: 
\begin{eqnarray}
i \partial_t u_d + \frac{1}{2} \partial_{x}^2u_d  
-(|u_d|^2 + g_{12}|u_b|^2 -1) u_d &=& 0,
\label{deq11}
\\
i \partial_t u_b +\frac{1}{2} \partial_{x}^2u_b  
-(g_{12}|u_d|^2 + g_{22}|u_b|^2- \mu) u_b &=& 0,
\label{deq21}
\end{eqnarray}
where 
$\mu\equiv\mu_b/\mu_d$ is the rescaled chemical potential.

The above system of equations conserves the total energy:
\begin{eqnarray}
E &=& \frac{1}{2}\int_{-\infty}^{+\infty} \mathcal{E} dx, \nonumber \\
\mathcal{E} &=& |\partial_{x} u_d|^2+|\partial_{x} u_b|^2+(|u_d|^2-1)^2
+g_{22}|u_b|^4 \nonumber \\
&-&2\mu |u_b|^2 + 2g_{12} |u_d|^2 |u_b|^2,
\label{energy}
\end{eqnarray}
as well as the number of atoms in each component
$N_d$ and $N_b$ and the total number of atoms, 
$N=N_d+N_b=\sum_{i=d,b}\int^{\infty}_{-\infty} dx|u_i|^2$. 
Furthermore, in the special case where the nonlinear coefficients are all 
equal to each other, i.e. $g_{12}=g_{22}=1$, Eqs.~(\ref{deq11})-(\ref{deq21}) correspond to the integrable Manakov model~\cite{Manakov}.
In such a case, the system admits exact single DB soliton solutions of the form:
\begin{eqnarray}
\!\!\!\!\!\!
u_d(x,t)&=&\left(\cos\phi\tanh\left[D(x-x_0(t))\right]+i\sin\phi\right),
\label{dark}
\\
\!\!\!\!\!\!
u_b(x,t)&=&\eta {\rm sech}\left[D(x-x_0(t))\right]\nonumber\\
&\times & \exp\left[ikx+i\theta(t)+i(\mu-1)t\right],
\label{bright}
\end{eqnarray}
subject to the boundary conditions $|u_d|^2\rightarrow 1$, and $|u_b|^2\rightarrow 0$ for $|x|\rightarrow  \infty$.
In the aforementioned solutions, $\phi$ is the so-called soliton's phase 
angle, while $\cos \phi$ and $\eta$ denote the amplitude of the dark and bright 
component respectively. Furthermore, $x_0(t)$ and $D$ correspond to the soliton's center 
and inverse width respectively, while $k=D \tan \phi$ is the constant 
wave-number of the bright soliton, associated with the
speed of the DB soliton, and $\theta(t)$ is its phase.

In the variational considerations that follow, we will utilize
Eqs.~(\ref{dark})-(\ref{bright}) as an ansatz for the general case where the interaction coefficients are
unequal. 
However, note in passing that substituting Eqs.~(\ref{dark})-(\ref{bright})
into the original system of Eqs.~(\ref{deq11})-(\ref{deq21}),
leads to certain conditions that the 
soliton parameters must satisfy for such a solution to exist
(as an exact solution). 
As shown in Ref.~\cite{ef}, the soliton 
parameters are connected via the following equations:
\begin{eqnarray}
D^2&=& \cos^2 \phi - g_{12} \eta^2,
\label{width1} \\[1.0ex]
D^2&=& g_{12}\cos^2 \phi - g_{22} \eta^2,
\label{width2} \\[1.0ex]
\dot{x}_0 &=& D\tan\phi,
\label{x0} \\
\theta(t)&=&\frac{1}{2}(D^2-k^2)t+(1-g_{12})t,
\label{omegat}
\end{eqnarray}
where $\dot{x}_0=d x_0/dt$ is the DB soliton velocity, together with the 
closure conditions regarding the width and amplitude of the solitons:
$\eta^2 =\cos^2 \phi \left(g_{12}-1\right)/\left(g_{22}-g_{12}\right)$, and
$D^2 = \cos^2 \phi \left(g_{22}-g^2_{12}\right)/\left(g_{22}-g_{12}\right).$
 We also note that in the rescaled system, the amplitude 
$\eta$ of the bright soliton, 
as well as the inverse width
parameter $D$ of the DB-soliton are 
connected to the number of atoms of the bright component by means of the 
following equation: 
$N_b\equiv \int |u_b|^2dx=2\eta^2/D.$  
It is important to highlight that we will not rely on
Eqs.~(\ref{width1})--(\ref{omegat}) for the analytical considerations
that follow. This is because these ``restrictive'' special solutions
only represent particular members of the family of DB solutions
that is also parametrically restricted by the conditions of
positivity of $\eta^2$ and $D^2$ in the above expressions.
Here we would like to consider the interaction coefficients and, in principle,
also the chemical potentials as free parameters, widely varying over
both the miscible and the immiscible regime. Then, generally, exact DB solitons 
following the profile of Eqs.~(\ref{dark})-(\ref{bright}) do not exist, but numerically
we can identify similar DB states that deviate only slightly from the tanh-sech shape over a wide range of model parameters.

\subsection{Interactions of two dark-bright solitons for general nonlinear coefficients}
In what follows, we will attempt to generalize the findings of 
Refs.~\cite{pe3,ef} by considering the interaction of two DB
solitons in the more general
case of arbitrary interaction coefficients, thus going beyond the integrable limit.
To describe a two DB soliton state 
in the absence of a confining potential, we will use as initial ansatz both 
for the analytical and the numerical considerations to be presented below,
a pair of two equal-amplitude single DB solitons travelling in 
opposite directions and having the form: 
\begin{eqnarray}
u_d(x,t)
&=& \left(\cos\phi\tanh X_{-}+i\sin\phi \right)
\nonumber \\ &\times &
\left(\cos\phi\tanh X_{+}-i\sin\phi\right),
\label{2dbd}
\\[1.0ex]
u_b(x,t)
&=& \eta\, {\rm sech} X_{-}\, {\rm e}^{i\left[kx+\theta(t)
+(\mu-1)t
\right]}
\nonumber \\&+&
\eta\, {\rm sech} X_{+}\, {\rm e}^{i\left[-kx+\theta(t)
+(\mu-1)t
\right]}\,
{\rm e}^{i\Delta\theta}.
\label{2dbb}
\end{eqnarray}
Here $X_{\pm} = D\left(x \pm x_0(t)\right)$, $2x_0$ is the relative 
distance between the two DB solitons, while $\Delta\theta$ is the 
relative phase between the bright solitons. 
Below we will consider both the IP, $\Delta\theta=0$, and OP, $\Delta\theta=\pi$, case. 
As noted above, using a tanh-sech profile for the individual DB soliton is in itself an approximation away from the integrable 
limit. Its validity will be discussed further in the next section.

In what follows, we will employ a Hamiltonian variational approach,
where the ansatz of Eqs.~(\ref{2dbd})-(\ref{2dbb}) is substituted
into the energy of Eq.~(\ref{energy}). Furthermore, and so as to perform the relevant integrations,
we assume that the soliton's velocity is sufficiently small, thus  $\cos(kx)\approx1$, and $\sin(kx)\approx0$. 
The final result for the total energy reads: $E = 2E_1+ E_{\rm dd}+ E_{\rm bb} + E_{\rm db}$ using the notation of
Ref.~\cite{pe3} for a direct comparison of the results.
 Introducing $\chi=4\eta^2/D$ (satisfying $\chi \approx N_b$ if the bright solitons are sufficiently separated), 
the different contributions to the energy are
given by the following expressions: 
\begin{eqnarray}
E_1 &=&\frac{4}{3}D^3 + \frac{1}{6}\chi D^2 \left(2g_{12} +3\tan^2\phi +1\right)\nonumber\\
&+& \frac{1}{6}\chi^2 D \left(g_{22}-
g^2_{12}\right)+\chi \left(g_{12}-\mu \right),
\label{e1} \\
E_{\rm dd}&\approx & 16 \cos^2\phi \left[ \frac{1}{3}D \cos^2\phi +D +2(\cos^2\phi -D^2) x_0 
\right.\nonumber \\
&-&\left.\frac{3+4\cos^2\phi}{3D} \cos^2\phi \right]{\rm e}^{-4Dx_0},
\label{edd}
\\[3.0ex]
E_{\rm bb}&\approx &\chi \Big[ 2D \left(D\left(1-Dx_0\right) -k^2 x_0\right) 
+g_{22} D \chi \Big]\nonumber \\
&\times& \cos\Delta\theta {\rm e}^{-2Dx_0} \nonumber \\
&+&g_{22} D\chi^2 \left(2Dx_0-1\right)\left(1+2\cos^2\Delta\theta\right){\rm 
e}^{-4Dx_0},
\label{ebb}
\\[3.0ex]
E_{\rm db}&\approx &-4\chi \left[ D x_0\left(\mu - g_{12}\right)  
+g_{12}\cos^2\phi\right]\times  \cos\Delta\theta {\rm e}^{-2Dx_0} \nonumber 
\\
&+&g_{12}\chi \cos^2\phi \Bigg[\frac{16}{3}\cos^2\phi-16Dx_0+8 \Bigg]{\rm 
e}^{-4Dx_0}.
\label{edb}
\end{eqnarray}
The terms in the aforementioned equations correspond to: the energy of a single 
DB soliton $E_1$ (contributing twice to the total energy), a result compatible with the one 
found in the very recent work of 
Ref.~\cite{ef}, and the interaction energies between the two dark, the two bright and the dark and bright 
solitons denoted as $E_{dd}$, $E_{bb}$ and $E_{db}$ respectively.
It is important to note here that the above expressions,
similarly to~\cite{pe3}, capture the dominant contributions of
the different energies thereby removing higher order terms (e.g.,
proportional to $e^{-6 D x_0}$ and higher). This is an assumption
that we will relax below.

By a direct comparison of the above results with the ones stemming from the integrable limit of the 
theory, obtained in Ref.~\cite{pe3}, it is immediately evident that
the interaction energy between the two dark solitons
is identical to the expression of~\cite{pe3}, a feature which is
expected given our effective rescaling of the interaction coefficients
by $g_{11}$.
In particular, also by this comparison, it becomes evident that 
the difference in the total energy of the system for 
general $g_{ij}$ comes from the bright solitons hosted in the second component 
of the coupled GPEs, Eq.~(\ref{deq21}). 
Notice the pre-factors $g_{22}$ entering in both terms of Eq.~(\ref{ebb}), 
and $g_{12}$ appearing in all terms of Eq.~(\ref{edb}).
Furthermore, and even more importantly, an extra term enters in 
the interaction energy, $E_{db}$, between the DB solitons, i.e. the 
first term appearing in Eq.~(\ref{edb}). Such a term,
which was not accounted for in the respective integrable limit of the 
theory studied in Ref.~\cite{pe3}, will significantly contribute in 
the final expression for the forces acting between the two DB 
solitons. 
Furthermore, and since this term is a leading order 
contribution to the interaction energy,
it directly suggests depending on its sign,
that the possibility may exist of identifying bound states 
{\it even for IP bright solitons} for general nonlinear coefficients. 

An important improvement of the results of~\cite{pe3}, in addition
to evaluating the relevant formulae for general $g_{ij}$ is
that we have also been able to {\it analytically integrate}
the relevant expressions i.e., to obtain the ``exact'' rather
than approximate integral results. The resulting energy
forms are as follows:
\begin{figure}[tbp]
\includegraphics[scale=0.32]{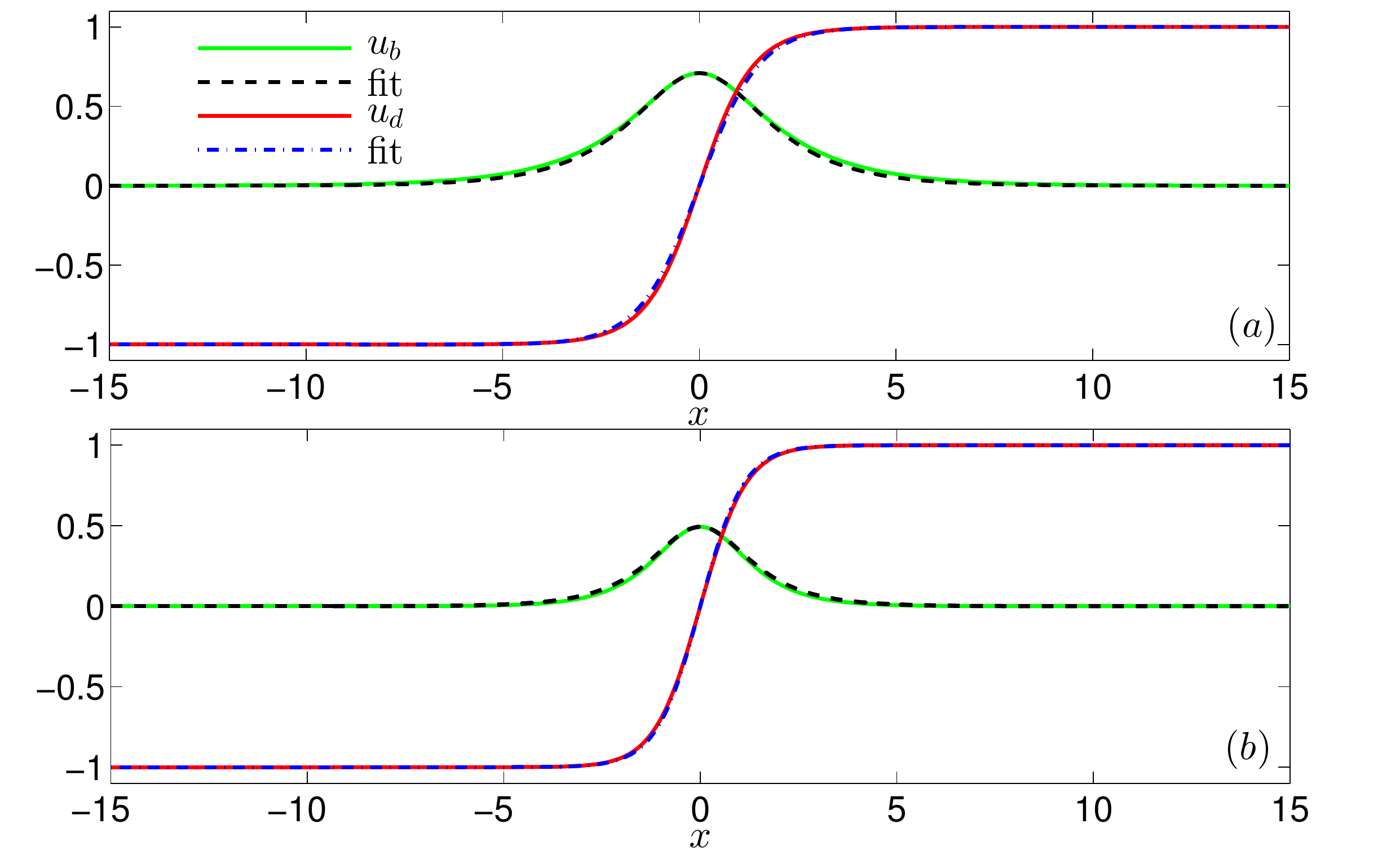}
\caption{(Color online): Profiles of the fitted single DB solitons, where $u_d$ and $u_b$ denoted by solid lines
respectively correspond to the analytical solution for the single DB case, while dashed and dashed-dotted lines 
refer to the fitted profiles for the bright and the dark soliton respectively, with (a) $g_{12}=0.85$, and (b) 
$g_{12}=1.2$. Notice the very good agreement of the fitted solutions upon passing from the miscible to the immiscible regime. }
\label{figfit}
\end{figure}

 %
 \begin{widetext}
\begin{eqnarray}
E_{\rm dd}&= &-\frac{11}{3D} \cos^8 \phi \csch^7 (2Dx_0) \sinh(6Dx_0)
+ \frac{4}{D} \cos^4 \phi \csch^2 (2Dx_0)\Big[-1+2Dx_0 \coth(2Dx_0)\Big]\nonumber \\
&+& \frac{1}{3D} \cos^8 \phi \csch^7 (2Dx_0) \Big[ 12Dx_0\left( 9 \cosh (2Dx_0)+ \cosh (6Dx_0)\right)-27\sinh (2Dx_0) \Big]\nonumber\\
&+&\frac{1}{3} D \cos^4 \phi \csch^5 (2Dx_0) 
\Big[-24 D x_0 \cosh (2Dx_0) +9 \sinh (2Dx_0) + \sinh (6Dx_0) \Big]  \nonumber \\
&-&\frac{4}{3D} \cos^6 \phi \csch^5 (2Dx_0)\Big[-24Dx_0\cosh(2Dx_0)+9 \sinh (2Dx_0)+ \sinh (6Dx_0) \Big] \nonumber \\
&+&\frac{1}{3} D \cos^2 \phi \Big[ -24 D x_0 \csch^3 (2Dx_0) \cosh (2Dx_0)+12 \csch^2 (2Dx_0)\Big], 
\label{Edd2}
\\[3.0ex]
E_{\rm bb}&=&-4 \eta^2 k^2x_0 \csch (2Dx_0)\cos\Delta\theta
 + 2\eta^2 D \csch^3(2Dx_0)\cos\Delta\theta  \Big[- Dx_0 \left[3+\cosh(4Dx_0)\right]\Big]\nonumber \\
&+&\frac{4}{D}g_{22}\eta^4\csch^3(2Dx_0)\cos\Delta\theta \Big[-4Dx_0+\sinh(4Dx_0) \Big]-8 \mu \eta^2 x_0\csch(2Dx_0)\cos\Delta\theta \nonumber \\
&+&\frac{4}{D}g_{22}\eta^4\csch^2(2Dx_0)\left( 1+2\cos^2 \Delta\theta \right)\Big[ -1+2Dx_0\coth(2Dx_0)\Big]\nonumber \\
&+& 2\eta^2 D \csch^3(2Dx_0)\cos\Delta\theta \sinh(4D x_0), 
\label{Ebb2}
\\[3.0ex]
E_{\rm db}&=& 8g_{12}\eta^2 x_0 \csch(2Dx_0)\cos\Delta\theta -\frac{8}{D}g_{12}\eta^2 \cos^2 \phi \csch^2(2Dx_0)\Big[-1+2Dx_0 
\coth(2Dx_0)\Big] \nonumber \\
&+&\frac{4}{3D}g_{12}\eta^2 \cos^4 \phi \csch^5 (2Dx_0)\Big[ -24Dx_0 \cosh(2Dx_0) +9\sinh(2Dx_0) +\sinh(6Dx_0)\Big]\nonumber \\
&+& \frac{4}{D}g_{12}\eta^2 \cos^4 \phi \csch^5 (2Dx_0)\cos\Delta\theta \Big[ 4Dx_0\left[2+\cosh(4Dx_0) \right]-3\sinh(4Dx_0)\Big] \nonumber \\
&-&\frac{4}{D}g_{12}\eta^2 \cos^2 \phi \csch^3 (2Dx_0) \cos\Delta\theta \Big[-4Dx_0 +\sinh(4Dx_0)\Big]. 
\label{Edb2}
\end{eqnarray}
\end{widetext}
 From these equations, the asymptotic results of Eqs.~(\ref{edd})-(\ref{edb}) can be recovered by expanding with respect to 
 $\exp(-2D x_0)$.
 A key realization is that both the bright-bright soliton energy
 $E_{bb}$
 and the cross-component interaction energy
$E_{db}$
 depend on $\cos(\Delta \theta)$,
i.e., on the relative phase and their contribution appears
{\it at the same order}. Hence, the key underlying intuition of
OP bright solitons yielding attraction that will 
counteract the repulsion of the dark solitons enabling
the existence of equilibria may not
be sufficient, in terms of providing quantitatively accurate results. 
The significant contribution of the interaction
of the dark solitary waves of one component with the bright ones of
the other must be factored in.
On the other hand, we also see why, at least in principle,
a bound state is possible. The different asymptotic rates of decay of the dark
soliton interaction, i.e. faster decaying as $\exp(-4D x_0)$ and dominant at shorter distances  
[see also Eq.~(\ref{edd})],
with the bright (and DB) interaction, more slowly decaying as $\exp(-2D x_0)$ and
dominant at longer distances [see Eqs.~(\ref{ebb})-(\ref{edb})], may create the possibility of an
equilibrium, especially for the OP bright soliton scenario.
\begin{figure}[tbp]
\includegraphics[trim=0 0 0 0,clip,scale=0.45]{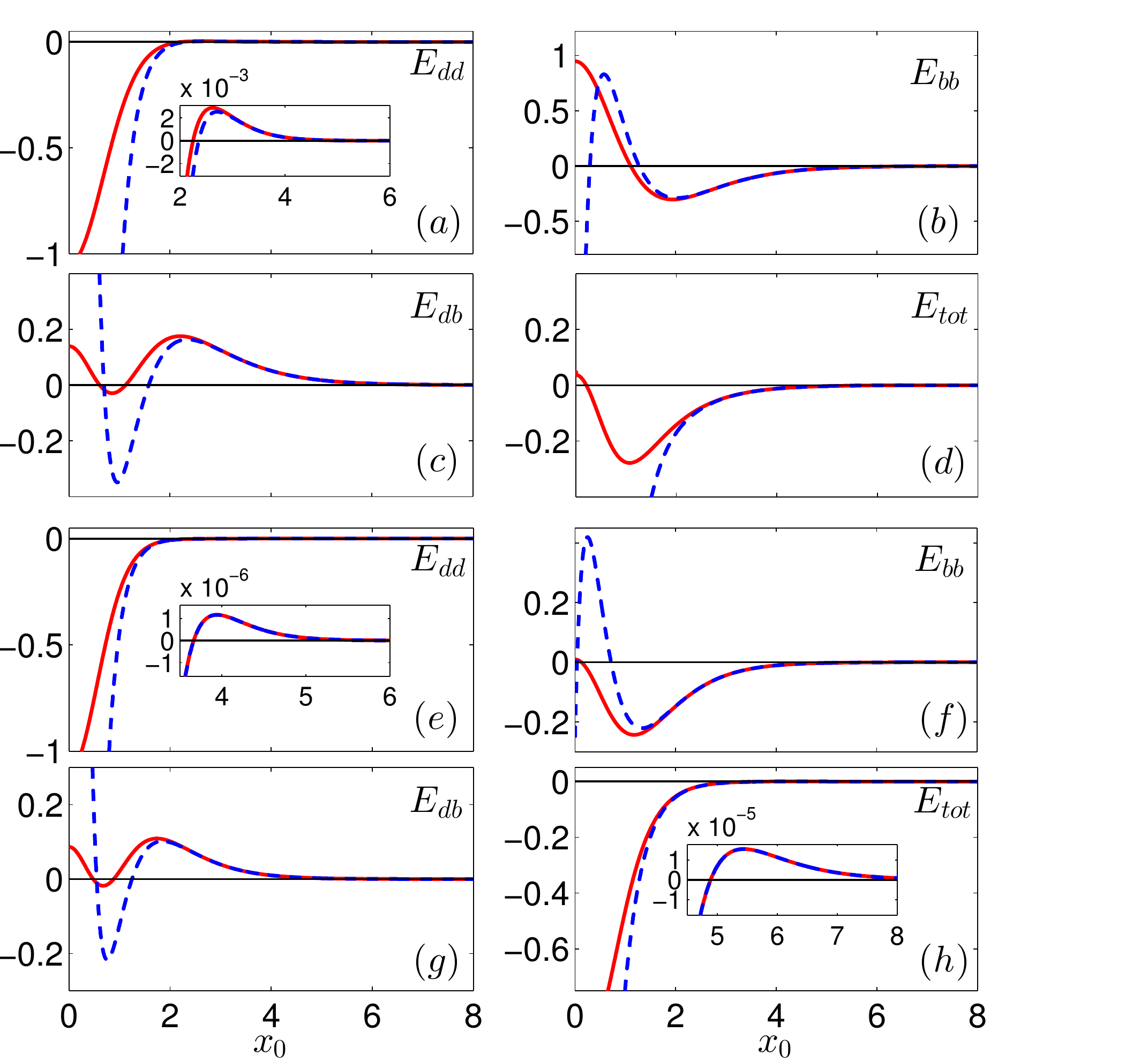}
\caption{(Color online): Comparison between the exact (solid red lines) and the approximate (dashed blue lines)
expressions, within the Hamiltonian variational formulation, for the individual interaction energies [see 
Eqs.~(\ref{Edd2})-(\ref{Edb2}) and Eqs.~(\ref{edd})-(\ref{edb})] 
for IP bright soliton components. From top left to bottom right shown are $E_{dd}$, $E_{bb}$, $E_{db}$ and $E_{tot}$ 
respectively. $(a)$-$(d)$ and $(e)$-$(h)$ correspond to $g_{12}=0.85$ and $g_{12}=1.2$, respectively. 
In all cases the inset provides a magnified version of the respective extremum. }
\label{fig3}
\end{figure}
\begin{figure}[tbp]
\includegraphics[scale=0.45]{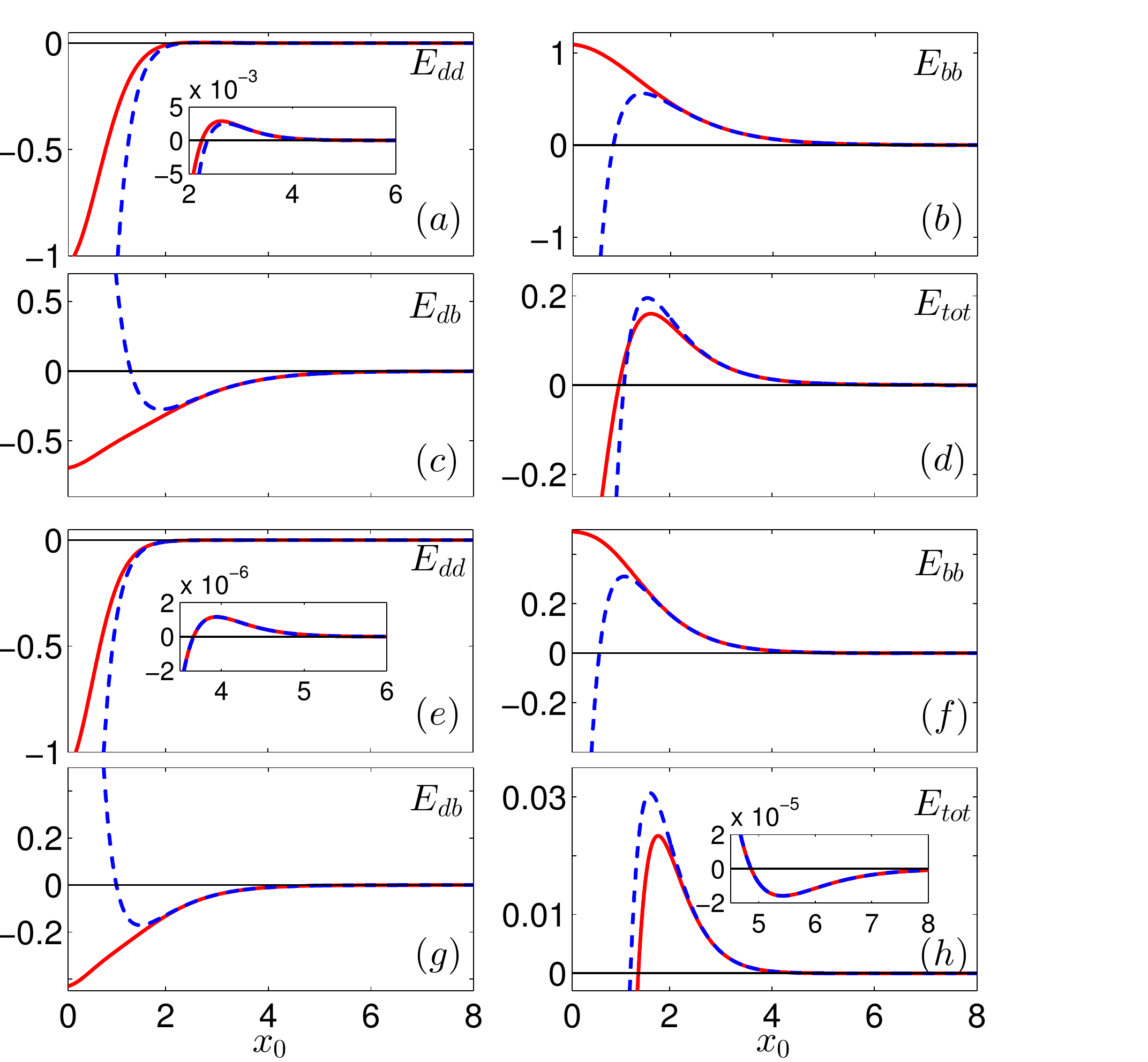}
\caption{(Color online): Same as Fig.~\ref{fig3} but for OP bright soliton components. }
\label{fig4}
\end{figure}

\section{Numerical Results}

Before we embark into an examination  of the energies for the
different values of the parameters, it is relevant to address the
issue of how good the fit of the single DB is with respect
to the analytically
available expression of Eqs.~(\ref{dark})-(\ref{bright}). Recall,
once again, that the expressions of Eqs.~(\ref{width1})-(\ref{omegat})
are not used, hence the inverse width parameter $D$ and
the bright soliton amplitude parameter $\eta$ are obtained
from the exact numerical --up to the prescribed accuracy-- single soliton solution upon fitting.
As shown in Fig.~\ref{figfit}, and in
line with the findings of~\cite{ef}, the analytical expression
[see Eqs.~(\ref{dark})-(\ref{bright})] is in very good agreement with
the numerically obtained solution
for $g_{12}$ in the immiscible regime [see Fig.~\ref{figfit} $(b)$], and especially as $g_{12}$
becomes large. On the other hand, in the miscible regime for
$g_{12}$, the tendency of the components to overlap 
deteriorates
the quality of the approximation, especially for $g_{12} \ll 1$ [see Fig.~\ref{figfit} $(a)$].
We remind the reader that the miscibility-immiscibility threshold
(after rescaling) is given by $g_{12}^2 = g_{22}$~\cite{aochui}. Hence, we expect
our analytical approximation to be progressively more accurate
as we go deeper into the immiscible regime and to be least adequate
for ratios of inter- to intra-component interactions well below unity.
Moreover, the improved quality of the fit is also assisted by the fact that upon increasing $g_{12}$ 
while keeping $\mu$ fixed, a steady decrease of the bright amplitude is observed (cf. also Fig.~{\ref{figbdg}}$(i)$ below). 
This acts in favor of the tanh-sech ansatz as it brings the system closer to the (integrable again) dark-only single-
component limit.

We now turn to an examination of the different energy contributions
for the IP case in Fig.~\ref{fig3} and for the OP
case in Fig.~\ref{fig4}. 
Focusing on stationary solutions, we choose $\phi=0$, $k=0$ and $\Delta \theta=0$ (IP) or $\Delta \theta=\pi$ (OP).
The chemical potential ratio is fixed to $\mu=\frac{2}{3}$. At varying $g_{12}$ we numerically identify the single-DB profile 
and extract the effective values of $D$ and $\eta$ 
by fitting the tanh-sech ansatz to it.
As in most of the cases that will follow,
we provide a representative case example in the miscible regime
with $g_{12}=0.85$ and another in the immiscible regime with
$g_{12}=1.2$. The remaining rescaled nonlinearity coefficient has been fixed to $g_{22}=0.95$, motivated by the relevant 
ratios in the case of $^{87}$Rb. Starting with the case of the dark-dark
soliton interaction 
[Fig.~\ref{fig3} $(a)$, $(e)$], we observe that it looks ``attractive''
in the sense that a regular particle would tend towards the
center, upon the imposition of such a potential energy landscape.
Yet, the negative effective mass of the DB solution~\cite{frantz} should be factored
in and in that case, indeed the interpretation is one of repulsion,
as expected from the above. The top right and bottom left panels
of each quartet of panels in Fig.~\ref{fig3}, represent respectively 
$E_{bb}$ [Fig.~\ref{fig3} $(b)$, $(f)$], and $E_{db}$ [Fig.~\ref{fig3} $(c)$, $(g)$]  confirming their
{\it opposite} trend as well as their comparable value.
In the IP case, at large distances, the bright-bright interaction is repulsive, while the
DB is attractive.
In the results shown,
both the {\it exact}, Eqs.~(\ref{Edd2})-(\ref{Edb2}), and
the {\it asymptotic},  Eqs.~(\ref{edd})-(\ref{edb}), forms of the energy
expressions are given (denoted by solid red and dashed blue lines 
respectively in Figs.~\ref{fig3},~{\ref{fig4}). 
Expectedly, they coincide at large $x_0$ where $\exp(-2Dx_0)$ becomes small, while at short distances the deviations can 
be substantial, even qualitatively.
It should be borne in mind here that {\it neither} of these expressions is sufficiently
good when the DB solitons are sufficiently close (distances $0< x_0 < 1.5$
typically for our results herein). There, the solitons are essentially
overlapping and hence, the superposition ansatz of Eqs.~(\ref{2dbd})-(\ref{2dbb}) clearly fails.

Coming to the main conclusion of the theoretical analysis, let us
examine the OP case of Fig.~\ref{fig4}. The key
panel to consider is that of the total energy [Fig.~\ref{fig4} $(d)$, $(h)$].
The identification
of a local maximum there for both the cases of $g_{12}=0.85$ and
$g_{12}=1.2$ (and for all values of $g_{12}$ in between)
suggests the existence of an effective local minimum representing a stable
equilibrium, at least in as far as the OP vibration
of the two DBs around it is concerned. Note here that due to the translational
invariance, 
the two-DB center of mass can always be displaced freely. 
This prediction can be tested against the direct numerical computations of the
original PDE system of Eqs.~(\ref{deq11})--(\ref{deq21}). In the latter,
we use a fixed point (Newton-Raphson) iteration to obtain the
OP equilibrium \cite{kelley}. Subsequently, we identify the location
of the (coincident between the two components) soliton center
and compare it to the corresponding theoretical value. The
result is shown in Fig.~\ref{fig2} and is quite satisfactory
in terms of capturing the relevant trend. The increasing tendency
of the equilibrium $x_0$ as a function of $g_{12}$ as the miscibility
threshold is crossed and as we move deeper into the immiscibility
regime can be intuitively understood as follows. In the OP
case, the bright-bright interaction [Fig.~\ref{fig3} $(b)$, $(f)$] 
is attractive at the tails and the DB interaction
is repulsive [Fig.~\ref{fig3} $(c)$, $(g)$]. 
This becomes even more pronounced as $g_{12}$ increases
as can be seen from Eq.~(\ref{Edb2}) [see also Eq.~(\ref{edb})] and thus 
the equilibrium is expected to be found at larger distances due to this
stronger repulsion.
\begin{figure}[tbp]
\includegraphics[scale=0.4]{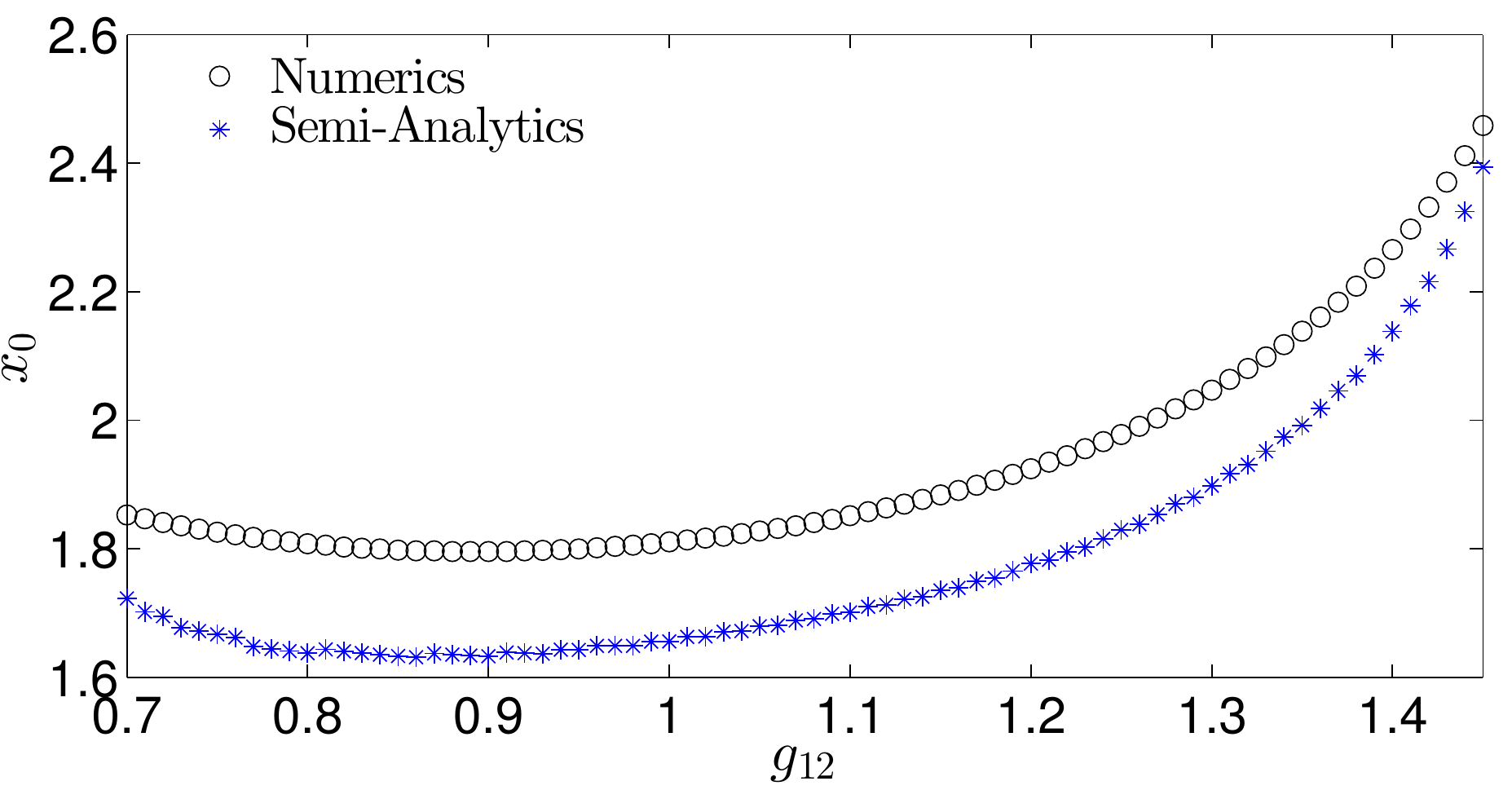}
\caption{(Color online): Equilibrium location $x_0$ as a function of the nonlinear coefficient $g_{12}$ for OP bright 
soliton components. Black circles depict the numerically obtained two DB soliton center, while
blue stars correspond to the semi-analytical prediction of the effective particle picture.}
\label{fig2}
\end{figure}
	
\begin{figure}[tbp]
\includegraphics[scale=0.4]{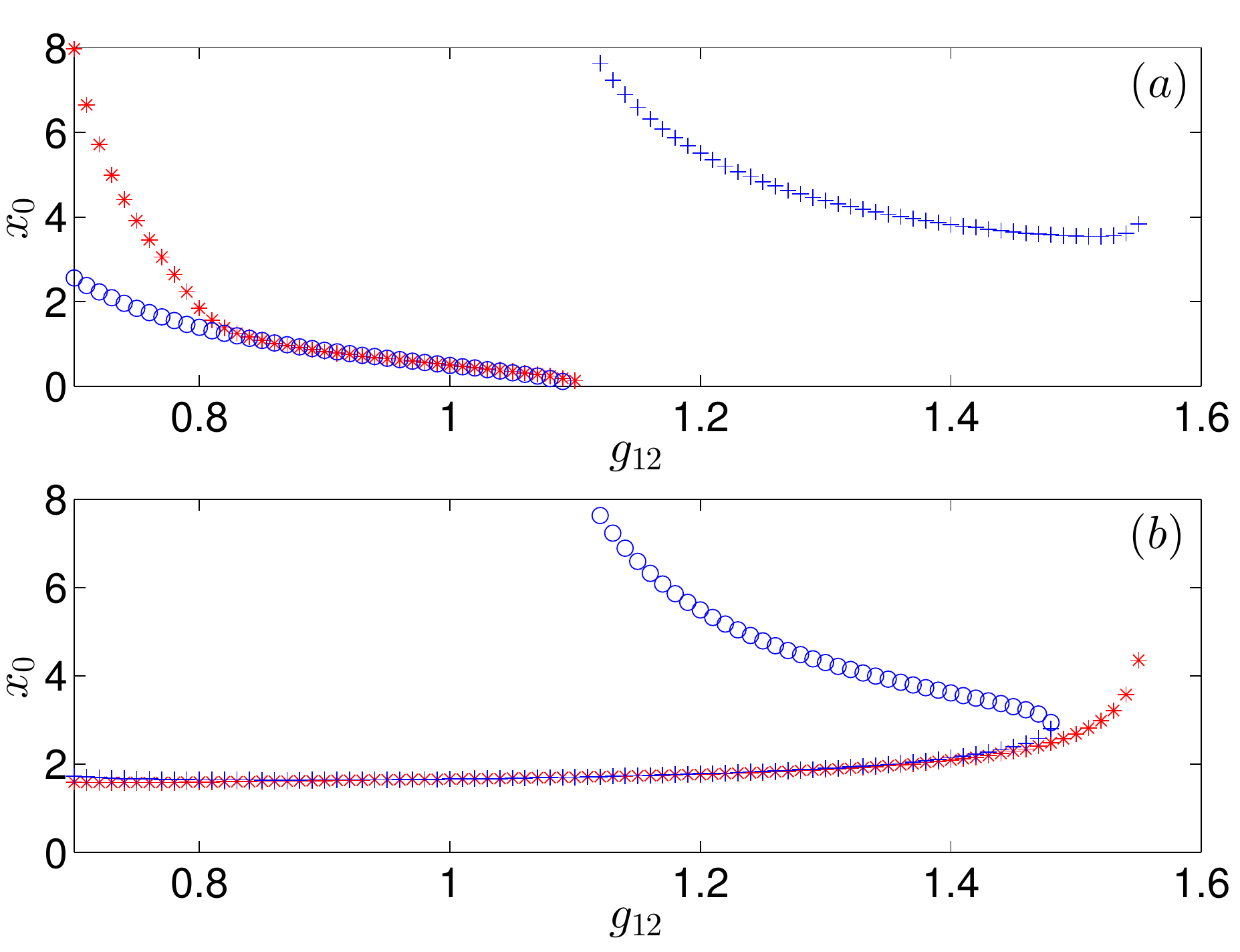}
\caption{(Color online): Variationally predicted equilibrium positions between two DB solitons
as a function of $g_{12}$. 
$(a)$ [$(b)$] shows the extrema for IP  (OP) bright soliton components.
Blue circles denote the location of the minimum of the total energy $E_{tot}$, 
while blue crosses correspond to 
the respective maxima. Red stars refer to the extrema predicted via the fully numerical evaluation of the energy (see text).
The chemical potential is fixed to $\mu=2/3$. }
\label{fig1}
\end{figure}
%

\begin{figure}[tbp]
\includegraphics[trim=0 0 0 0,clip,scale=0.44]{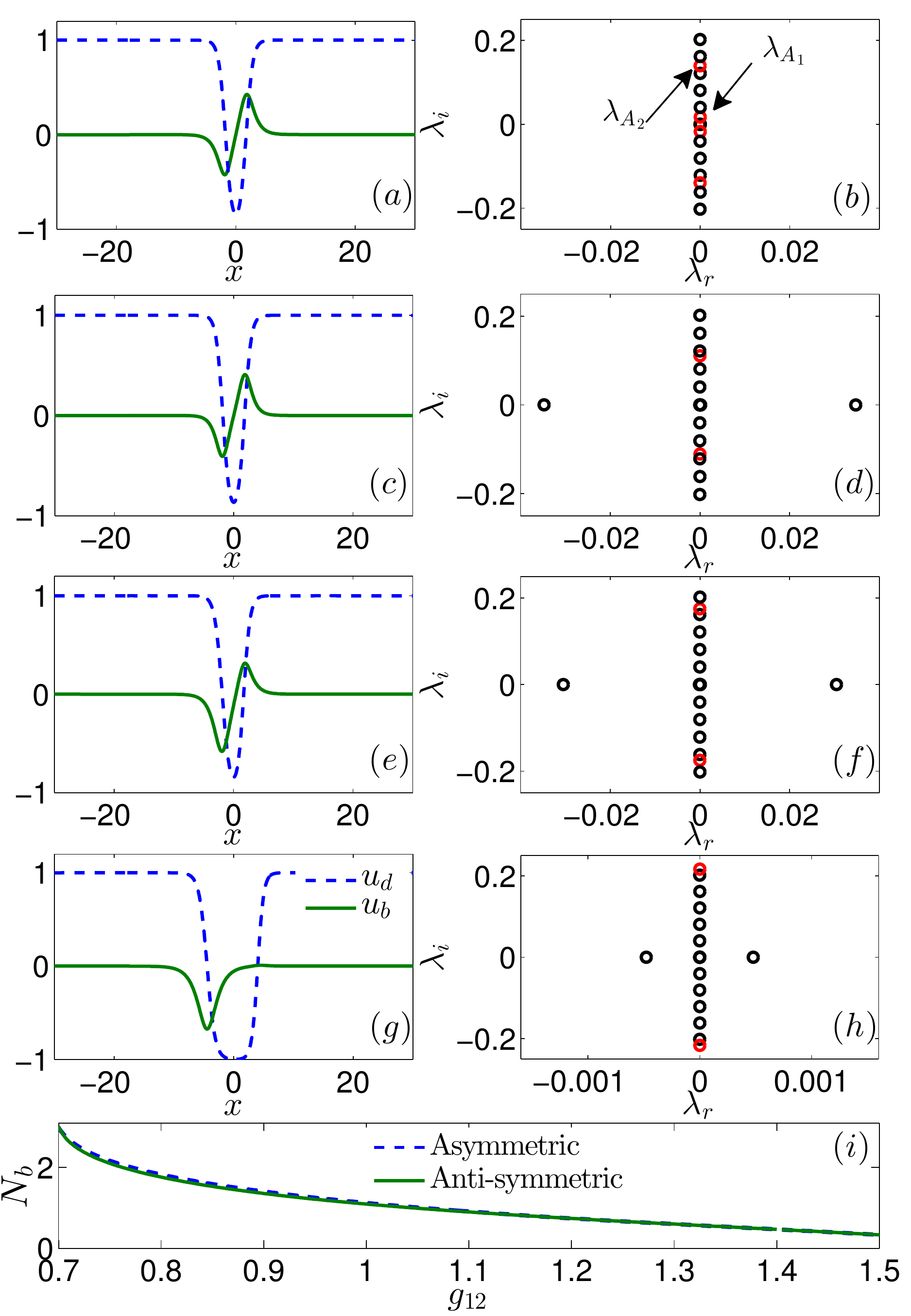}
\caption{(Color online): Left column: 
$(a)$ anti-symmetric stationary DB soliton pair for $g_{12}=1.1$, $(c)$ anti-symmetric state with $g_{12}=1.3$,
$(e)$ asymmetric state for $g_{12}=1.1$, and $(g)$ fully asymmetric D-DB state for $g_{12}=0.95$. 
Right columns [$(b)$ to $(h)$] are the associated BdG spectra, where the lowest twenty eigenvalues are shown with (black/red) 
circles. In all cases red circles denote the anomalous modes. $\lambda_{A_1}$, 
and $\lambda_{A_2}$, (indicated by black arrows) are the two modes related to the observed symmetry breaking, and 
the OP vibration of the 2-DB state respectively.
$(i)$: $N_b$ as a function of $g_{12}$ for the two different branches~\cite{sg}.} 
\label{figbdg}
\end{figure}

\begin{figure}[tbp]
\includegraphics[trim=0 0 0 0,clip,scale=0.44]{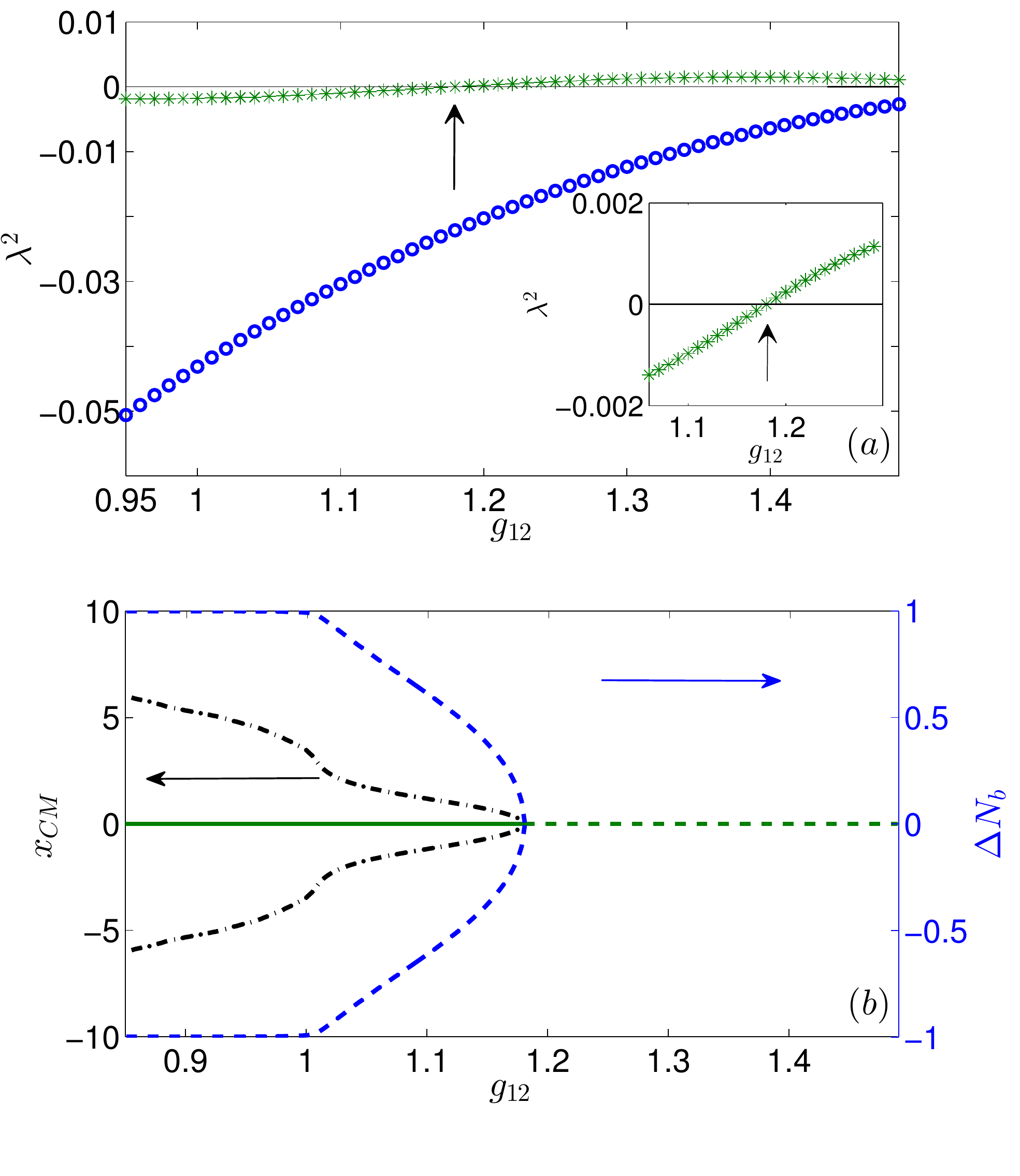}
\caption{(Color online): $(a)$ Trajectories of the
  squared eigenvalues of the
  two anomalous modes, $\lambda^2$ upon varying $g_{12}$ for the anti-symmetric stationary solutions of 
Eqs.~(\ref{deq11})-(\ref{deq21}). 
Blue circles correspond to the   
trajectory of the mode related to the OP vibration of the 2-DB state. 
Green stars denote the trajectory of the mode responsible for the symmetry breaking bifurcation that occurs  
at $g_{12_{cr}}=1.18$.
$(b)$ bifurcation diagram, obtained via measuring the center of mass and the mass imbalance of the bright component 
(see the text for the relevant definitions) as 
functions of $g_{12}$. The two arrows indicate the different axes used. 
Four branches are identified: three unstable ones (dashed-dotted black and dashed blue/green lines)  
and a stable branch (solid green line).  
Notice that the asymmetric branches exist before the critical point and are unstable, i.e. the pitchfork is subcritical.}  
\label{figbdg1}
\end{figure}

Now, we turn to a series of spurious features that the theory
may produce. By far the most significant is that if, by comparison,
we examine the total energy in the bottom right panels of 
the quartets in Fig.~\ref{fig3} $(d)$, $(h)$,
we will notice the existence of a local minimum, which is tantamount
to an effective energy maximum, and is expected to correspond
to a {\it saddle point} in the case of the IP bright solitons
(within the two DBs). This is found to exist only for sufficiently
small $g_{12}$'s i.e., it exists for $g_{12}=0.85$, but not for
$g_{12}=1.2$ (cf. $(a)$-$(d)$  vs. $(e)$-$(h)$ quartets of panels in Fig.~\ref{fig3}).
This feature, if present, would suggest that while in the IP
case the dark-dark and bright-bright interactions are both repulsive,
the DB one is attractive and enough to counter both to produce
an equilibrium, albeit an unstable one.
However, an extensive effort to identify this feature in the
PDE did not lead to fruition. While we cannot fully exclude the
possibility that such an equilibrium exists, both our fixed
point iteration results and the direct numerical simulations
(presented below) suggest its absence. Our explanation for
this ``negative'' result is that this equilibrium arises
at relatively short distances and rapidly moves towards the
origin. The relevant dependence on $g_{12}$ can be found
in Fig.~\ref{fig1} $(a)$ denoted by blue circles. Here, we observe
that the equilibrium occurs at moderate distances
only for the smallest values of $g_{12}$ considered
(close to $0.7$) where we know that the ansatz is the worst
(within our range) as regards capturing the true waveform
of the DB soliton. As $g_{12}$ increases, while the ansatz
gradually improves, the equilibrium distance rapidly decreases,
rendering the ansatz inadequate due to the overlap (and
constructive interference in this case) between the solitons.
Therefore, the individual character of the solitons is lost and
hence the ansatz again fails. Thus, we intend this part
as a cautionary tale about the potential inadequacies
of the variational ansatz, either due to the failure
of the profile of Eqs.~(\ref{dark})-(\ref{bright}) or because
of the failure of the two-soliton waveform.

An additional feature that we have identified 
when taking the Hamiltonian variational formulation at face
value can be seen in the insets of Figs.~\ref{fig3} and
\ref{fig4}, when looking at sufficiently
large distances in the case of $g_{12}=1.2$.
Examining the total energy plots at the bottom right panel of each 
of the aforementioned figures [Fig.~\ref{fig3} $(h)$ and 
Fig.~\ref{fig4} $(h)$],
we find that an additional extremum appears to arise. The trajectory
of this extremum (in both cases) is illustrated
in Fig.~\ref{fig1} $(a)$ (blue crosses) for the IP
and in Fig.~\ref{fig1} $(b)$ (blue circles) for the OP case.
A careful inspection of the relevant energy scales of
Figs.~\ref{fig3} and~\ref{fig4} confirms that this is a
{\it miniscule} effect of the order of $10^{-5}$, in this case example.
Again, an extensive search for corresponding stationary solutions on the PDE level
remained unsuccessful, which suggests that the miniscule large-$x_0$ extrema of the $E_{tot}$ energy curves
may be artifacts of the variational approach that are absent in the true dynamics. 

In fact, in an attempt to explore these spurious extrema further and to disentangle the errors 
stemming from the single-DB fitting process
and its concomitant identification of $D$ and $\eta$ on the one hand
and the two-DB ansatz in Eqs.~(\ref{2dbd})-(\ref{2dbb}) on the other hand,
we also attempted a ``purely numerical'' construction
of the energy. That is, upon identifying the single-DB waveform,
we numerically constructed the 2-DB profile, multiplying two
darks and adding (for the IP case) two brights
centered at varying distances,  i.e. in the spirit of Eqs.~(\ref{2dbd})-(\ref{2dbb}) but avoiding the tanh-sech fit.
Then, we numerically evaluated the resulting energy
by numerically integrating the expression of Eq.~(\ref{energy}).
This side-steps the inadequacies of the single soliton
ansatz, yet it does not avoid the issues of the two soliton
ansatz upon close proximity of the bright solitons.
 Within this latter approach, we do not find the miniscule large-$x_0$ maximum in $E_{tot}$ for the IP case,
suggesting that this is a spurious feature induced by the imperfect tanh-sech fit.
In contrast, the more substantial in-phase energy minimum for $g_{12}\lesssim 1.1$ qualitatively persists even without 
performing this fit [see red stars in Fig.~\ref{fig1} $(a	)$].
Its predicted quantitative position, however, is shifted to
considerably larger values of $x_0$ for small $g_{12}$.
This discrepancy highlights the inaccuracies of the fit in the miscible regime, but the {\it existence} of the spurious IP 
minimum itself seems to be induced not by the imperfect single-DB fit but by the construction of the two-DB ansatz.
Finally, in the OP case the fully numerical variational approach predicts a bond length of the DB pair that agrees well with 
that from the tanh-sech fit approach, underlining the robustness of this key result. The fit-based method also predicts a 
local minimum in $E_{tot}$ 
which annihilates the bound-state maximum in a saddle-center bifurcation at $g_{12}= 1.48$. Without the tanh-sech fit, the 
bound state is predicted
to persist up to $g_{12}=1.55$, beyond which the bright norm is zero [see red stars in Fig.~\ref{fig1} $(b)$]. 
In the numerical variational framework we do not see a 
minimum in $E_{tot}$ here, but possibly the aforementioned termination of the bound-pair branch could also
be caused by a saddle-center bifurcation happening at large $x_0$ (where all energies become extremely small and ultimately 
lie below our numerical resolution).

\begin{figure}[tbp]
\includegraphics[scale=0.45]{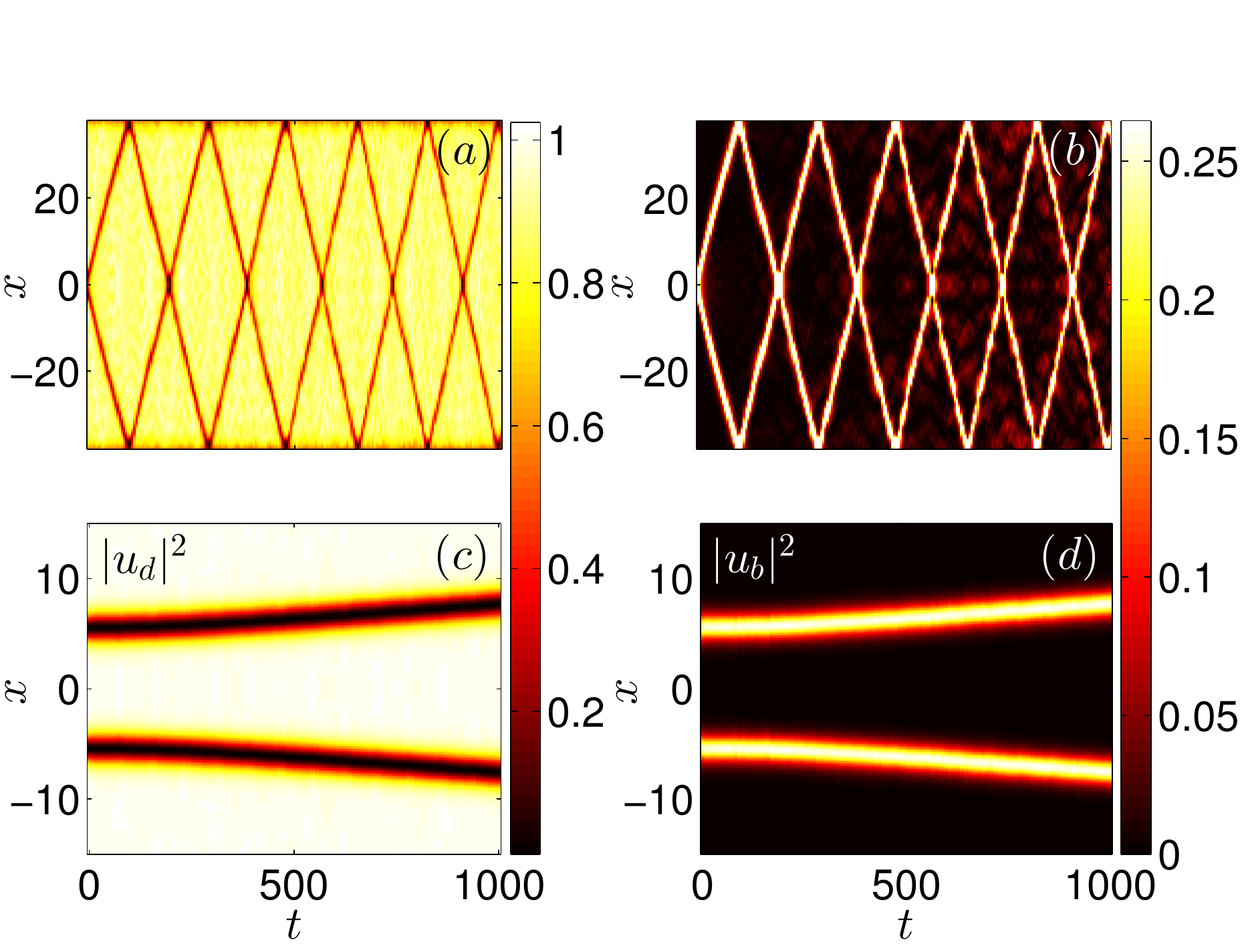}
\caption{(Color online): Typical examples of the space-time evolution of the density of a 2-DB soliton state 
for IP bright solitons. $(a)$, $(b)$ densities for $g_{12}=0.85$  
and equilibrium distance 
$x_0=1.37$. $(c)$, $(d)$ same as the above, but for $g_{12}=1.2$ and 
$x_0=5.45$.  In both cases  $(a)$, $(c)$ [$(b)$, $(d)$] 	 
refer to the evolution of the dark [bright] soliton component.} 
\label{fig5}
\end{figure}

It is then clear that out of the four possible extrema presented
in the in and out of phase cases, only the out of phase
equilibrium is relevant for the system. Hence, we explore
the latter further, before delving into the associated dynamics.
To assess the stability of this fixed point, we perform
a BdG analysis, linearizing around the
equilibrium as follows:
\begin{eqnarray}
  u_d &=& u_d^{(eq)} + \left(a(x) e^{-i \omega t} + b^{\star}(x) e^{i \omega^{\star} t}
  \right),
  \label{bdg1}
  \\
 u_b &=& u_b^{(eq)} + \left(c(x) e^{-i \omega t} + d^{\star}(x) e^{i \omega^{\star} t}
  \right). 
  \label{bdg2}
\end{eqnarray}
The resulting linearization system for the eigenfrequencies $\omega$
(or equivalently eigenvalues $\lambda=i \omega$) and
eigenfunctions $(a,b,c,d)^T$ is solved numerically. If modes with purely
real eigenvalues (genuinely imaginary eigenfrequencies) or complex
eigenvalues (eigenfrequencies) are identified, these are tantamount
to the existence of an instability~\cite{revip}. Remarkably, indeed
this is the case, as we increase $g_{12}$. An eigenvalue pair
crosses the origin and becomes real, resulting in an exponential
instability that we will trace soon in the dynamics as well.
Presently, we inquire which mode could be responsible for such
an instability. We note that in addition to 6 modes in the spectrum
at $\lambda=\omega=0$ due to symmetries (the conservations of
$N_d$ and $N_b$, as well as due to the momentum-conservation-inducing
invariance with respect to translation), there are two additional
modes of interest that are ``hidden'' within the continuous spectrum
of the problem, see the BdG spectrum in the middle right panel of Fig.~\ref{figbdg}
where black arrows indicate the modes in question denoted by red circles. 
These modes are so-called anomalous or negative energy
modes. They possess  
negative energy or negative Krein
signature~\cite{skryabin}. The mode energy (or Krein signature)   is
defined as
\begin{eqnarray}
  K= \omega \int  \Big(|a|^2 - |b|^2 + |c|^2 - |d|^2\Big) dx,
  \label{krein}
\end{eqnarray}
in a multi-component system like the one considered herein.
Our computations show that there are {\it two} such modes
in the system. One is the anticipated one, related to the
out-of-phase vibration of the two DB solitons ($\lambda_{A_2}$ in Fig.~\ref{figbdg} $(b)$). 
However, as suggested by the energetics discussed previously it remains
stable (pertaining to oscillations in the effective
energy landscape discussed above). The second mode ($\lambda_{A_1}$ in Fig.~\ref{figbdg} $(b)$)
is the one associated with {\it symmetry breaking} of the
bright component. I.e., adding the corresponding eigenvector
to the bright component breaks its symmetry and results
in two bright solitons of unequal amplitudes.
It is the latter mode that destabilizes at the instability
threshold of the out of phase configuration.
The existence (and destabilization)
of such a mode suggests the presence of a pitchfork bifurcation
associated with the symmetry breaking. Indeed, we have been
able to identify the asymmetric branches related to this
pitchfork bifurcation. Interestingly, the branches exist
before the critical point of the anti-symmetric solution
destabilization and are unstable themselves. That is to say the pitchfork is {\it subcritical}.
A similar, yet crucially different, bifurcation mechanism was identified in the work of~\cite{et}
in the presence of a parabolic trap.
In that setting, an effective
single-component description was found to be applicable, where the two dark solitons 
together with the trap act as an approximately static double-well 
potential for the bright component. Within a two-node tight-binding 
approximation this then maps to the bosonic Josephson junction model which features a symmetry-breaking 
{\it supercritical} pitchfork bifurcation from the OP branch.
Correspondingly, the emerging asymmetric modes are found to be stable in the trapped framework,
while they are unstable in our present setting [see the linearization spectra in Figs.~\ref{figbdg}$(f,h)$], highlighting 
the fundamentally different nature of the symmetry-breaking pitchfork bifurcations in the trapped and untrapped cases,
respectively.

In Fig.~{\ref{figbdg}} $(a)$, an anti-symmetric solution is shown  
together with its linearization spectrum for $g_{12}=1.1$ [Fig.~{\ref{figbdg}} $(b)$].
Two anomalous modes are identified, 
indicated by the arrows. 
The lowest of these eigenvalues, namely $\lambda_{A_1}$, moves towards zero with increasing $g_{12}$
and eventually crosses to the real axis, which signals the pitchfork bifurcation.
In panels $(c)$-$(d)$, an example of an anti-symmetric equilibrium state is illustrated past its destabilization threshold.  
Furthermore, in panels $(e)$-$(h)$ asymmetric solutions are identified 
and shown together with their respective linearization 
eigenvalues manifesting their instability. In particular, $(e)$-$(f)$ correspond to an asymmetric
solution for $g_{12}=1.1$, providing a straightforward comparison of its BdG spectrum with the respective spectrum of 
the anti-symmetric state in $(a)$-$(b)$. Notice in particular the absence of the anomalous mode $\lambda_{A_1}$.
Moreover, in panels $(g)$-$(h)$ another example of the asymmetric state at smaller $g_{12}$ is depicted. Here, a total 
transfer of mass of the 
bright component to one of the dark solitons has occurred, resulting in a bound state of one purely dark and one dark-bright 
soliton.
Note again the crucial difference to the self-trapped states of the bosonic Josephson junction here, since in the present 
setting the pitchfork bifurcation phenomenology arises in a spatially homogeneous setting.
Finally, Fig.~{\ref{figbdg}}$(i)$ illustrates the decrease of the bright soliton norm upon increasing $g_{12}$
for both the anti-symmetric and asymmetric solitonic states. 

All of the above findings are summarized in Fig.~{\ref{figbdg1}}. 
In particular, Fig.~{\ref{figbdg1}} $(a)$ shows the trajectory of the
squared eigenvalues of the two anomalous modes as $g_{12}$ increases.
Notice that among these two trajectories the lowest one (blue circles) 
asymptotically tends to zero,
and as such it remains stable for all values of $g_{12}$. This trajectory corresponds to the OP vibration
of the two DB state.
However, a completely different picture is painted by the trajectory of the second mode 
(green stars). Closely following this trajectory [see also the inset in Fig.~\ref{figbdg1} $(a)$] 
we see that this mode destabilizes at $g_{12_{cr}}=1.18$ (which corresponds to the eigenvalue zero crossing). 
This destabilization signals the bifurcation and the emergence of the upper (black dotted) branch.
Note that for clarity reasons we only show values of $g_{12}$ in the vicinity of the bifurcation.              
The respective bifurcation diagram is shown in Fig.~{\ref{figbdg1}} $(b)$.
To obtain this diagram we simply measure the center of mass between the two bright solitons in the 
second component, i.e. 
$x_{CM}=\int_{-\infty}^{\infty}  x|u_b|^2 dx/\int^{\infty}_{-\infty}|u_b|^2dx$,
upon varying $g_{12}$.
This way, we can identify both the stable anti-symmetric branch [solid green line in Fig.~\ref{figbdg1} $(b)$]
as well as the three unstable branches, two asymmetric and one anti-symmetric
[see dashed-dotted black lines and dashed-dotted green line respectively in Fig.~\ref{figbdg1} $(b)$].
Notice that the asymmetric branches exist before the critical point,   
verifying the subcritical pitchfork nature of the relevant bifurcation. 
It is also worth mentioning the ``neck'' in $x_{CM}$ that occurs at the immiscibility to miscibility transition.
This feature is found to coincide with a change in the character of the symmetry-broken soliton configuration,
i.e. the transition from a gradually asymmetric DB-DB pair as in Fig.~{\ref{figbdg}} $(e)$ to a maximally asymmetric 
dark/dark-bright (D-DB) state as in Fig.~{\ref{figbdg}}$(g)$.
This is seen in the relative bright imbalance defined as $\Delta N_b\equiv\left(\int_{-\infty}^0 |u_b|^2 dx-\int^{\infty}_0 |
u_b|^2 dx\right)/\int^{\infty}_{-\infty}|u_b|^2dx$ and shown in dashed blue lines in Fig.~\ref{figbdg1}$(b)$.

\begin{figure}[tbp]
\includegraphics[scale=0.45]{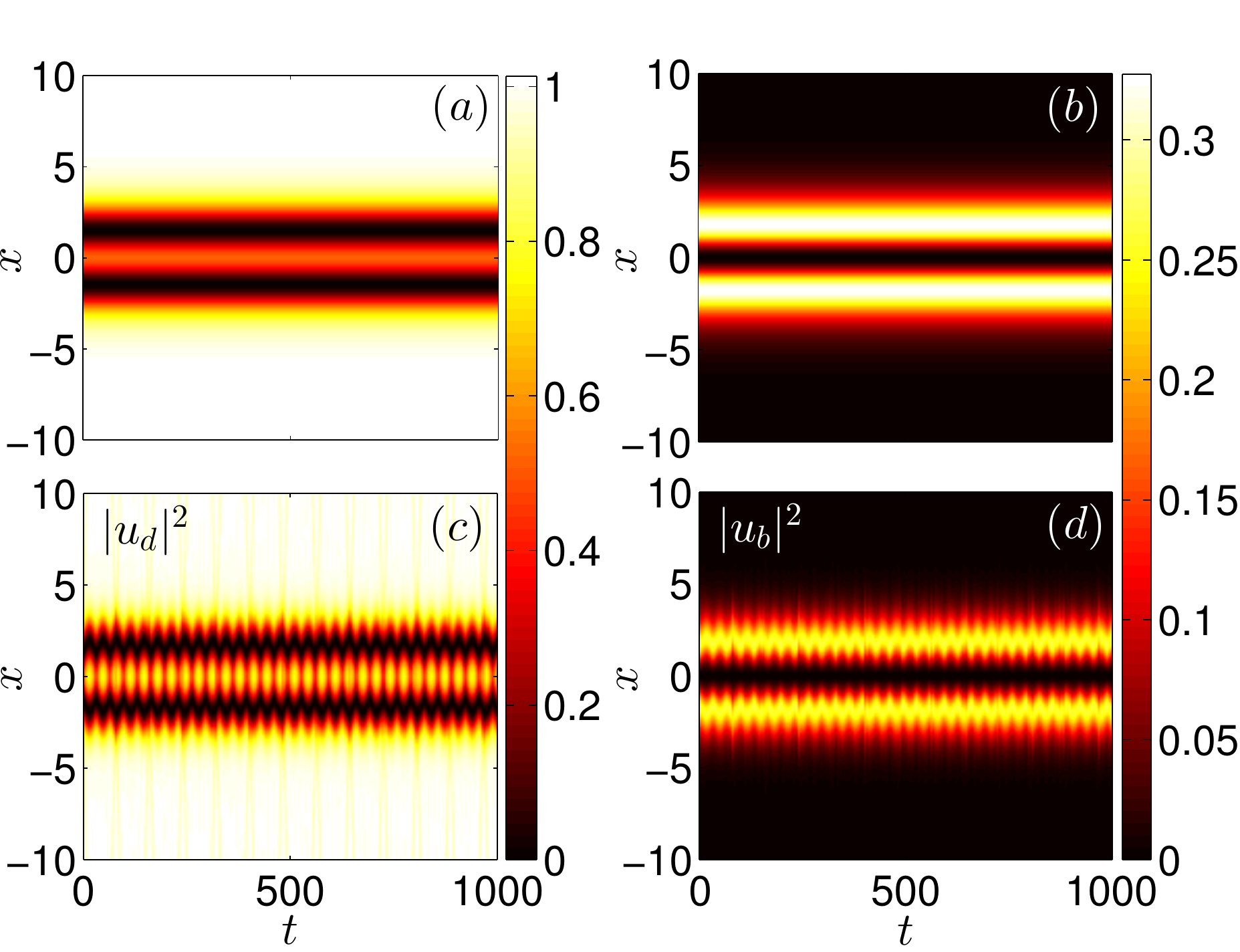}
\caption{(Color online): Typical examples of the space-time evolution of the density of a stationary 2-DB soliton state 
for OP bright solitons and $g_{12}=0.85$. 
Evolution of the dark soliton density $(a)$,  
and the respective bright one $(b)$, 
initialized at the predicted equilibrium distance with $x_0=1.65$. 
$(c)$, $(d)$ same as the above, but slightly inside 
of the predicted fixed point at $x_1=1.45$.} 
\label{fig6}
\end{figure}


Finally, we turn to direct numerical simulations corroborating
the existence and stability results presented above.
Firstly, in Fig.~\ref{fig5}, we examine the (theoretically predicted,
yet argued as spurious) equilibria of the theoretical energy
analysis presented above. We see in the figure that both in the
case of $g_{12}=0.85$ [Fig.~\ref{fig5} $(a)$, $(b)$] and in that of $g_{12}=1.2$ 
[Fig.~\ref{fig5} $(c)$, $(d)$], as well
as in all the additional cases that we have examined (not shown
here), repulsive dynamics is manifested between the two DBs.
We have tried different distances, smaller as well as larger
than the equilibrium one, always finding this type of repulsive
behaviour. If the prediction of a saddle point was an accurate
one, the solitons should have moved inward (rather than outward,
i.e., featuring attraction) when initially positioned below
the equilibrium distance. In fact, what we find is that at
short distances the soliton repulsion is fairly dramatic.
As the figure suggests, this is partially the result of
constructive interference in the case of our initial conditions.
I.e., while adjusting from an initial profile through a transient
stage, the solitary waves emit radiative wavepackets. Some
of these move outward (and do not pose concerns, provided
that the integration domain is sufficiently large). However,  
some of these move inward, constructively interfere and exert
upon return a substantial effective force which assists the solitons
towards initiating their outward trajectory.
Thus, our direct simulations also confirm the expectation that
the variationally predicted fixed point is an artifact of the
ansatz and is dynamically absent in the IP case.

\begin{figure}[tbp]
\includegraphics[trim=0 0 0 0,clip,scale=0.45]{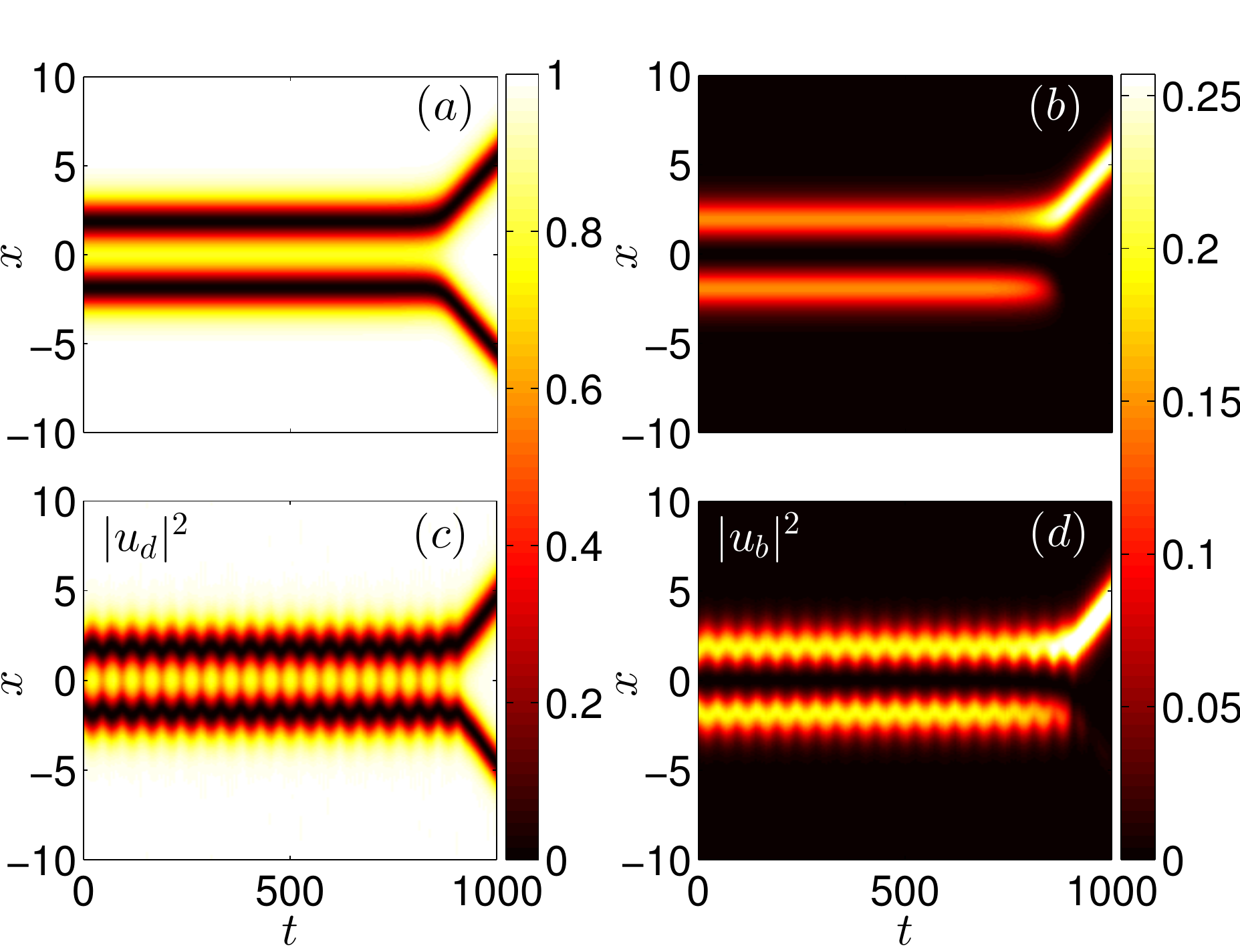}
\caption{(Color online): Same as Fig.~\ref{fig6} but for $g_{12}=1.2$, i.e.
well inside the immiscible regime. 
$(a)$, $(b)$ initialization at $x_0=1.78$, coinciding with the semi-analytical 
prediction of the stationary 2-DB state. 
$(c)$, $(d)$ initialization at $x_1=1.58$. Notice that symmetry breaking 
occurs towards the end of the simulation, resulting in an almost complete transfer 
of mass from the left bright soliton to the 
right one [$(b)$, $(d)$ counterpart of each doublet].}
\label{fig7}
\end{figure}

Hence, we hereafter focus on the OP case. We explore the
latter both when initiated {\it at} equilibrium, as well as when
initiated {\it near} equilibrium. This is done for
the case of $g_{12}=0.85$ in Fig.~\ref{fig6}, as well as for
that of $g_{12}=1.2$ in Fig.~\ref{fig7}. In the former case,
our BdG stability analysis predicts that the anti-symmetric
bright soliton configuration (and the whole DB pair) will
be dynamically robust. Indeed, this is what we observe;
when initializing at the equilibrium, [Fig.~\ref{fig6} $(a)$, $(b)$] the solitary waves
stay put, while when initiating at slightly smaller or larger
distances [Fig.~\ref{fig6} $(c)$, $(d)$], we simply excite the stable OP vibrational
mode of the two-soliton molecule. However, a fundamentally
different picture is shown in Fig.~\ref{fig7} for the
case of $g_{12}=1.2$. Here, while for a long time the configuration
appears to be quiescent eventually for times $t>800$, the
bright soliton redistributes its mass dramatically [see Fig.~\ref{fig7} $(b)$], resulting
in a strong repulsion between the ensuing single DB (with a much
larger bright soliton mass) and the dark soliton (respectively,
stripped of its soliton mass, Fig.~\ref{fig7} $(a)$). This leads to the strong separation
of the solitary waves as a result of the dynamics. The same feature
can be seen in the case of oscillations around the equilibrium, Fig.~\ref{fig7} $(c)$, $(d)$;
while the solitons appear robust for many oscillation cycles,
we can see them eventually redistributing the bright component [Fig.~\ref{fig7} $(d)$]
and splitting as a result. Although the instability appears to
be dramatic and instantaneous, a more careful monitoring of the
system suggests otherwise. In particular in Fig.~{\ref{fig8},
we monitor the evolution of the bright density at the central point between the
two solitons. Examining the relevant diagnostic in a semilog scale, we clearly infer
that the instability is building over the entire horizon
of the simulation, featuring a remarkable exponential growth
over many orders of magnitude until eventually it produces an
effect of order unity resulting in the symmetry breaking. We
confirm by examining the growth rate of this exponential growth,
that it is indeed occurring with the unstable eigenvalue of
the two-DB state.

\begin{figure}[tbp]
\includegraphics[trim=0 0 0 0,clip,scale=0.45]{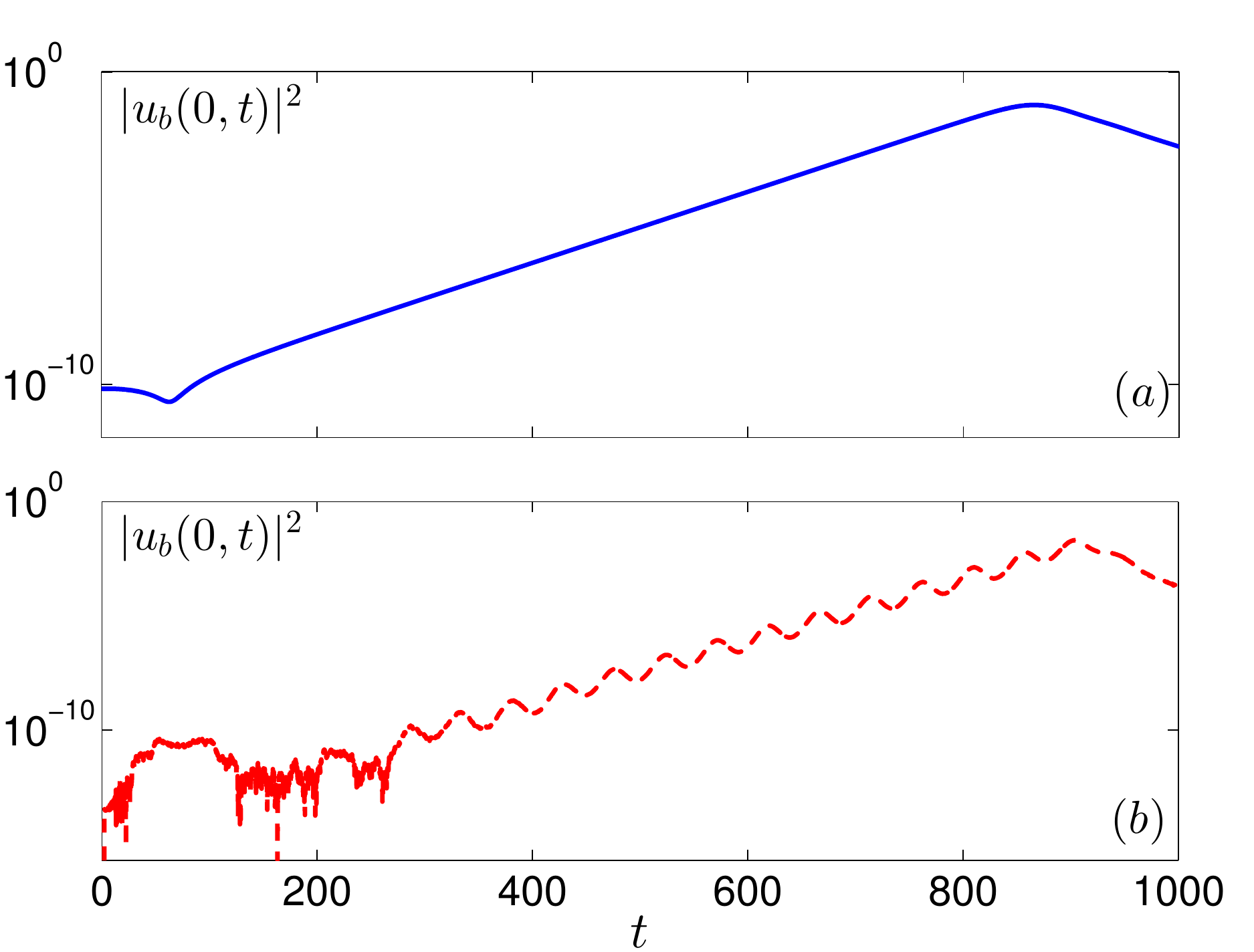}
\caption{(Color online): Semi-logarithmic representation, showing the central bright density between the two solitons
throughout the propagation depicted in  
Fig.~\ref{fig7}. This way, we monitor the exponential growth of the instability, 
that clearly  builds up from the first instant (initially hidden in the noise below our numerical accuracy threshold) 
and fully manifests itself when it becomes of order $\mathcal{O}(1)$ 
Panels $(a)$ and $(b)$ correspond to the symmetry breaking observed in panels $(b)$ and $(d)$  
of Fig.~\ref{fig7}.}
\label{fig8}
\end{figure}

\section{Conclusions and Future Challenges}

In the present work, the intriguing problem of
DB soliton interactions has been revisited. Motivated by recent experimental
studies of the problem, the relevant formulation has been extended
in a number of ways. We have considered the effect of general (and
beyond integrable) inter-atomic interaction coefficients and 
identified full analytical expressions of the variational energy-based
formulation, rather than solely approximate ones, focusing
on the former as being more accurate than the latter.
Our aim was to explore the conclusions of the energy based calculation
monoparametrically varying the inter-component scattering length.
This led to the identification of key additional features in the energy
such as the significant role of DB soliton
interaction, often overlooked in earlier studies. We carefully
considered which of the features of the variational formulation
are credible for leading to accurate results and which ones should
be discarded and for what reason i.e., which assumptions and approximations
in the variational formulation may turn out to fail. Focusing
on the predominant nontrivial feature, namely the existence of
an equilibrium in the out-of-phase case, we showcased the
predictive strength of the formulation and identified a
subcritical pitchfork bifurcation instability-inducing
scenario that had not been
previously observed, to the best of our knowledge. The consequences
of the instability were dynamically explored and observed
to lead to the key phenomenon of mass redistribution.

There is a multitude of intriguing questions that
are worth examining in future efforts. On the one hand,
it would be particularly interesting to explore
if the variational formulation might predict the potential
for symmetry-breaking instability provided the relevant freedom.
More concretely, we can utilize an ansatz for the bright solitons
involving two hyperbolic secants with distinct amplitudes $\eta_1$
and $\eta_2$. An important question is: is this sufficient
(as one might hope/expect based on the above description)
to observe the instability at the level of the few degree of
freedom system? In the context of this symmetry-breaking, 
it would moreover be interesting to explore the crossover from our present untrapped setting
with its subcritical pitchfork bifurcation to the harmonically trapped case
where in \cite{et} a supercritical pitchfork bifurcation has been found to destabilize the out-of-phase mode.
Extensions of the present considerations
would also be worthwhile to pursue in other settings
including ones involving a higher number of components, as
well as higher dimensions. In the former one,
solitary waves such as dark-dark-bright and
dark-bright-bright ones~\cite{DDB} have been predicted,
so it would be interesting to see how the relevant phenomenology
generalizes. In the latter one, the role of dark solitons is played by
vortices~\cite{kody,pola}.
Such ``vortex-bright'' solitons are quite robust and furthermore their
vorticity is topologically protected, so it would be interesting
to examine whether they would form similar bound states and
what the stability and dynamics of the latter would be.
Studies along these directions are reserved to future works.

\section*{Acknowledgements}
P. S. gratefully acknowledges financial support by the 
Deutsche Forschungsgemeinschaft (DFG)
in the framework of the grant SCHM 885/26-1.
P.G.K. gratefully acknowledges the
support of NSF-DMS- 1312856, NSF-PHY-1602994, the
Alexander von Humboldt Foundation, and the ERC under
FP7, Marie Curie Actions, People, International Research
Staff Exchange Scheme (IRSES-605096).


\begin{thebibliography}{99}

\bibitem{revip} P.G. Kevrekidis, D.J. Frantzeskakis, Reviews in Physics {\bf 1}, 140 (2016).

  \bibitem{yuri1} Yu. S. Kivshar and G. P. Agrawal, {\it Optical solitons: from fibers to photonic crystals}
(Academic Press, San Diego, 2003).


\bibitem{christo} D. N. Christodoulides,
Phys. Lett. A \textbf{132}, 451 (1988).

\bibitem{vdbysk1} 
V. V. Afanasjev, Yu. S. Kivshar, V. V. Konotop, and V. N. Serkin, 
Opt. Lett. \textbf{14}, 805 (1989).

\bibitem{vddyuri} 
Yu. S. Kivshar and S. K. Turitsyn, 
Opt.\ Lett. \textbf{18}, 337 (1993).

\bibitem{ralak} 
R. Radhakrishnan and M. Lakshmanan, 
J.\ Phys.\ A: Math.\ Gen. \textbf{28}, 2683 (1995).

\bibitem{dbysk2} 
A. V. Buryak, Yu. S. Kivshar, and D. F. Parker, 
Phys. Lett. A \textbf{215}, 57 (1996).

\bibitem{shepkiv} 
A. P. Sheppard and Yu. S. Kivshar, 
Phys.\ Rev.\ E \textbf{55}, 4773 (1997).

\bibitem{parkshin} 
 Park Q. Han, and Shin H. J.,
Phys.\ Rev.\ E \textbf{61}, 3093 (2000).
	
\bibitem{seg1} Z. Chen, M. Segev, T. H. Coskun, D. N. Christodoulides, and
Yu. S. Kivshar, 
J. Opt. Soc. Am. B \textbf{14}, 3066 (1997).

\bibitem{seg2} 
E. A. Ostrovskaya, Yu. S. Kivshar, Z. Chen, and M. Segev, 
Opt.\ Lett. \textbf{24}, 327 (1999).

\bibitem{BA} Th. Busch, and J. R. Anglin, Phys. Rev. Lett. {\bf 87}, 010401 (2001).


\bibitem{hamburg} 
C. Becker, S. Stellmer, P. Soltan-Panahi, S. D{\"o}rscher, M. Baumert, E.-M.
Richter, J. Kronj\"{a}ger, K. Bongs, and K. Sengstock, 
Nat. Phys. \textbf{4}, 496 (2008).

\bibitem{pe1} 
C. Hamner, J. J. Chang, P. Engels, and M.A. Hoefer, 
Phys.\ Rev.\ Lett. \textbf{106}, 065302 (2011).

\bibitem{pe2} 
S. Middelkamp, J. J. Chang, C. Hamner, R. Carretero-Gonz{\'{a}}lez, P. G.
Kevrekidis, V. Achilleos, D. J. Frantzeskakis, P. Schmelcher, and P. Engels,
Phys.\ Lett.\ A \textbf{375}, 642 (2011).

\bibitem{pe3} 
D. Yan, J. J. Chang, C. Hamner, P. G. Kevrekidis, P. Engels, V. Achilleos,
D. J. Frantzeskakis, R. Carretero-Gonz{\'{a}}lez, and P. Schmelcher, 
Phys.\ Rev.\ A \textbf{84}, 053630 (2011).

\bibitem{azu} 
A. {\'{A}}lvarez, J. Cuevas, F. R.. Romero, C. Hamner, J. J. Chang, P.
Engels, P. G. Kevrekidis, and D. J. Frantzeskakis, 
J. Phys. B 
\textbf{46}, 065302 (2013).


\bibitem{pe4} 
M.A. Hoefer, J. J. Chang, C. Hamner, and P. Engels, 
Phys.\ Rev.\ A \textbf{84}, 041605(R) (2011).

\bibitem{pe5} 
D. Yan, J. J. Chang, C. Hamner, M. Hoefer, P. G. Kevrekidis, P. Engels, V.
Achilleos, D. J. Frantzeskakis, and J. Cuevas, 
J.\ Phys.\ B: At.\ Mol.\ Opt.\ Phys. \textbf{45}, 115301 (2012).

\bibitem{krolkiv} Y. S. Kivshar and W. Kro´likowski, Opt. Commun. {\bf 114},
353 (1995).

\bibitem{markus1} A. Weller, J. P. Ronzheimer, C. Gross, J. Esteve, M. K. Oberthaler, D. J. Frantzeskakis, G. Theocharis, and 
P. G. Kevrekidis, Phys. Rev. Lett. {\bf 101}, 130401 (2008);
G. Theocharis, A. Weller, J. P. Ronzheimer, C. Gross, M. K. Oberthaler, P. G. Kevrekidis, and D. J. Frantzeskakis,
Phys. Rev. A {\bf 81}, 063604 (2010).

\bibitem{borisrh}  J.H.V. Nguyen, P. Dyke, D. Luo, B.A. Malomed,
  R.G. Hulet, Nat. Phys. {\bf 10}, 918 (2014).
  
\bibitem{fr1}  S. Inouye, M. R. Andrews, J. Stenger, H.-J. Miesner D. M. Stamper-Kurn, and W. Ketterle, Nature (London)
{\bf 392}, 151 (1998);  J. L. Roberts, N. R. Claussen, J. P. Burke, Jr., C. H. Greene, E. A. Cornell, 
and C. E. Wieman, Phys. Rev. Lett. {\bf 81}, 5109 (1998); 
E. A. Donley, N. R. Claussen, S. L. Cornish, J. L. Roberts, E. A. Cornell, and C. E. Wieman, Nature (London)
{\bf 412}, 295 (2001).

\bibitem{fr2} G. Thalhammer, G. Barontini, L. De Sarlo, J. Catani, F. Minardi, and M. Inguscio, 
Phys. Rev. Lett. {\bf 100}, 210402 (2008); S. B. Papp, J. M. Pino, and C. E. Wieman, Phys. Rev. Lett. {\bf 101}, 040402 (2008).
\bibitem{chin}
C. Chin, R. Grimm, P. Julienne, and E. Tiesinga,
Rev. Mod. Phys. {\bf 82}, 1225 (2010).

\bibitem{halll} 
K. M. Mertes, J. W. Merrill, R. Carretero-Gonz{\'{a}}lez, D. J. Frantzeskakis, P. G. Kevrekidis, and D. S. Hall, Phys. Rev. Lett. {\bf 99}, 190402 (2007). 

\bibitem{opanchuk} Although we use these values of $g_{11}$ and
  $g_{22}$ as ``typical'', it is worthwhile to mention that the precise value
  of the coefficients is still under 
  active investigation, with the most accurate values known presently
   being those of M. Egorov, B. Opanchuk, P. Drummond, B. V. Hall, P.
Hannaford, and A. I. Sidorov, Phys. Rev. A {\bf 87}, 053614
(2013).

\bibitem{aochui} P. Ao and S. T. Chui
Phys. Rev. A {\bf 58}, 4836 (1998)

\bibitem{stringari}
L.~P.~Pitaevskii and S.~Stringari,
{\it Bose-Einstein Condensation.}
Oxford University Press (Oxford, 2003).

\bibitem{siambook} P.~G.~Kevrekidis,
D.~J.~Frantzeskakis, and R.~Carretero-Gonz{\'a}lez,
{\it The Defocusing Nonlinear Schr{\"o}dinger Equation},
SIAM (Philadelphia, 2015).

\bibitem{Manakov} S. V. Manakov, Zh. Eksp. Teor. Fiz. {\bf 65}, 505 (1973)
[Sov. Phys. JETP {\bf 38}, 248 (1974)].

\bibitem{kelley} C. T. Kelley, {\it Solving Nonlinear Equations with Newton's Method} (Society for Industrial and Applied Mathematics, Philadelphia, 1995).


\bibitem{ef} D. Yan, F. Tsitoura, P. G. Kevrekidis, and D. J. Frantzeskakis,
Phys. Rev. A {\bf 91}, 023619 (2015).


\bibitem{frantz}  D. J. Frantzeskakis, J. Phys. A {\bf 43}, 213001 (2010).

\bibitem{skryabin} Dmitry V. Skryabin,
Phys. Rev. A {\bf 63}, 013602 (2000).
  
\bibitem{sg} 
Note that, for the numerical implementation we use hard-wall-like boundary conditions, specifically a super-Gaussian potential of the form $V(x)=20\left(1-e^{\left(-x/45\right)^{24}}\right)$, 
which introduces an unavoidable discretization of the continuous spectrum. 
For some parameters we numerically observe collisions of linearization modes from this discretized continuous spectrum with the anomalous modes resulting in oscillatory instability windows;
these are likely to depend on the specific boundary conditions and are not explored in more detail here.

\bibitem{et} E. T. Karamatskos, J. Stockhofe, P. G. Kevrekidis, and P. Schmelcher, Phys. Rev. A {\bf 91}, 043637 (2015).

\bibitem{DDB} H. E. Nistazakis, D. J. Frantzeskakis, P. G. Kevrekidis, B. A. Malomed, and R. Carretero-Gonz{\'a}lez,
Phys. Rev. A {\bf 77}, 033612 (2008).

\bibitem{kody} K. J. H. Law, P. G. Kevrekidis, and Laurette S. Tuckerman,
Phys. Rev. Lett. {\bf 105}, 160405 (2010).

\bibitem{pola} M. Pola, J. Stockhofe, P. Schmelcher, and P. G. Kevrekidis,
Phys. Rev. A {\bf 86}, 053601 (2012).



\end{thebibliography}
\end{document}